\documentclass[journal, comsoc]{IEEEtran}
\usepackage[letterpaper, left=0.625in, right=0.625in, top=0.75in, bottom=1.1in]{geometry}
\usepackage{amsmath,amsfonts}
\usepackage{algorithm}
\usepackage{array}
\usepackage[caption=false,font=footnotesize,labelfont=sf,textfont=sf]{subfig}
\usepackage{textcomp}
\usepackage{stfloats}
\usepackage{svg}
\usepackage{url}

\usepackage{verbatim}
\usepackage{graphicx}
\usepackage{siunitx}
\usepackage{booktabs}
\usepackage{xcolor}
\usepackage[export]{adjustbox}
\usepackage{cite}
\usepackage{algpseudocode}
\usepackage{multicol}
\usepackage{array}
\usepackage{colortbl}
\usepackage{tcolorbox}
\usepackage[hidelinks]{hyperref}
\usepackage{threeparttable}
\UseRawInputEncoding

\algdef{SE}
[STRUCT]
{Struct}
{EndStruct}
[1]
{\textbf{struct} \textsc{#1}}
{\textbf{end struct}}
\makeatletter
\newcounter{phase}[algorithm]
\newlength{\phaserulewidth}
\newcommand{\setphaserulewidth}{\setlength{\phaserulewidth}}

\newcommand{\phase}[1]{%
  \vspace{.25ex}
  \Statex\leavevmode\llap{\rule{\dimexpr\labelwidth+\labelsep}{\phaserulewidth}}\rule{\linewidth}{\phaserulewidth}
  \Statex\strut\refstepcounter{phase}\textit{#1}
  \vspace{-1.25ex}\Statex\leavevmode\llap{\rule{\dimexpr\labelwidth+\labelsep}{\phaserulewidth}}\rule{\linewidth}{\phaserulewidth}}
\makeatother

\setphaserulewidth{.7pt}

\NewDocumentCommand{\todo}{m o}{%
    \textcolor{red}{TODO-#1}%
}
\algnewcommand{\LineComment}[1]{\State \(\triangleright\) #1}

\newcommand\myatop[2]{\genfrac{}{}{0pt}{}{#1}{#2}} 
\newcommand{\dif}{\mathop{}\!\mathrm{d}} 

\begin{document}

\title{Routing and Spectrum Allocation in Broadband Quantum Entanglement Distribution}

\author{Rohan Bali,~\IEEEmembership{Student Member,~IEEE}, Ashley N. Tittelbaugh,~\IEEEmembership{Student Member,~IEEE}, Shelbi L. Jenkins,~\IEEEmembership{Member, ~IEEE},
Anuj Agrawal,~\IEEEmembership{Member,~IEEE,} Jerry Horgan,~\IEEEmembership{Senior Member,~IEEE}, Marco Ruffini,~\IEEEmembership{Senior Member,~IEEE}, Daniel C. Kilper,~\IEEEmembership{Senior Member,~IEEE}, Boulat A. Bash,~\IEEEmembership{Member,~IEEE}

\thanks{We acknowledge support from the National Science Foundation under Grant No. CNS-2107265 and Science Foundation Ireland grants 20/US/3708, 21/US-C2C/3750, and 13/RC/2077\_P2. This material is based upon High Performance Computing (HPC) resources supported by the University of Arizona TRIF, UITS, and Research, Innovation, and Impact (RII) and maintained by the UArizona Research Technologies department.}
\thanks{Some results from this manuscript were presented at the 2024 IEEE International Conference on Communications \cite{ICC_2024}.}
\thanks{Rohan Bali, Ashley N.~Tittelbaugh, Shelbi L.~Jenkins, and Boulat A.~Bash are with the Department of Electrical and Computer Engineering Department, University of Arizona, Tucson, AZ, USA.}
\thanks{Anuj Agrawal and Marco Ruffini are with the School of Computer Science and Statistics, CONNECT Center, Trinity College Dublin, Dublin, Ireland.}
\thanks{Jerry Horgan and Daniel C.~Kilper are with the Electronic and Electrical Engineering Department, CONNECT Centre, Trinity College Dublin, Dublin, Ireland.}
\thanks{Corresponding author: Rohan Bali, e-mail: \href{mailto:rbali@arizona.edu}{rbali@arizona.edu}.}
}

\markboth{Journal on Selected Areas in Communications,~Vol.~xx, No.~Y, August~2024}%
{Shell \MakeLowercase{\textit{et al.}}: A Sample Article Using IEEEtran.cls for IEEE Journals}

\IEEEpubid{0000--0000/00\$00.00~\copyright~2024 The Authors }

\maketitle
\thispagestyle{empty}

\begin{abstract}
We investigate resource allocation for quantum entanglement distribution over an optical network. 
We characterize and model a network architecture that employs a single broadband quasi-deterministic time-frequency heralded Einstein-Podolsky-Rosen (EPR) pair source, and develop a routing and spectrum allocation scheme for distributing entangled photon pairs over such a network. As our setting allows separately solving the routing and spectrum allocation problems, we first find an optimal polynomial-time routing algorithm. We then employ max-min fairness criterion for spectrum allocation, which presents an NP-hard problem. Thus, we focus on approximately-optimal schemes. We compare their performance by evaluating the max-min and median number of EPR-pair rates assigned by them, and the associated Jain index. We identify two polynomial-time approximation algorithms that perform well, or better than others under these metrics. We also investigate scalability by analyzing how the network size and connectivity affect performance using Watts-Strogatz random graphs. We find that a spectrum allocation approach that achieves higher minimum EPR-pair rate can perform significantly worse when the median EPR-pair rate, Jain index, and computational resources are considered. Additionally, we evaluate the effect of the source node placement on the performance.

\end{abstract}

\begin{IEEEkeywords}
quantum networks, optical fiber networks, quantum information science, routing protocols
\end{IEEEkeywords}

\section{Introduction}

\IEEEPARstart{Q}{uantum} entanglement resource distribution \cite{wehner2018quantum} can enable technologies like quantum-enhanced sensing \cite{Pirandola2018,quantumSensing}, distributed quantum computing \cite{distributedQuantumComputing}, and quantum security \cite{PhysRevLett.67.661}. The many entanglement distribution techniques can be categorized into repeaterless and quantum-repeater based, with their relative merits varying based on the scale and complexity of the networks, and their feasibility contingent on the advancements and future availability of new quantum devices. Here, we study the repeaterless approach, where photons travel along light-paths from the source of entanglement to the consumer node pairs. This is further sub-categorized into single- and multi-hop schemes.  The former directly connects the source to each consumer node via fiber links \cite{Wengerowsky,Zhu,Wang}, while the latter---the focus of this work---routes the photons to the consumer nodes via intermediate consumer nodes \cite{alshowkan2024}. In repeater-based techniques \cite{pant2019routing}, entanglement is routed to destination nodes via successive swaps of independently available entanglement resources on intermediate repeater nodes. 

While there has been extensive study of routing protocols in large-scale repeater networks \cite{Li2024Survey,Chen2024Fairness,Yang2022Fairness,Zhao2023mixed,muralidharan2016optimal,pant2019routing,wang2022pre,patil2022entanglement,panigrahy2023scalable,kaur2023distribution,sutcliffe2023multi,van2023entanglement}, the infancy of the quantum-repeater technology prohibits the near-term physical implementation of these networks. To date, work on repeaterless networks has focused on small-scale network demonstrations, thus overlooking routing and spectrum allocation methods that scale with network size. Our work addresses this gap by developing these for intermediate and large-scale repeaterless networks equipped with a single Einstein-Podolsky-Rosen (EPR)-pair source. \\
\IEEEpubidadjcol

Our work is motivated by the possibility of recently-proposed EPR-pair source that mitigates some of the loss limitations of repeaterless networks by generating time-frequency heralded entangled photons at both high rate and fidelity \cite{Chen}. Its detailed mathematical analysis is in \cite{shapiro2024entanglementsourcequantummemory}. This enables a `source-in-the-middle' entanglement-distribution network architecture, where the source node sends EPR pairs directly to two or more consumer nodes in the network, without the need for entanglement swapping. Heralding, or time-frequency stamping, of output EPR pairs, simplifies network management, entangled resource allocation, and quantum memory operation. However, properties of this and similar sources impose unique challenges in routing and spectrum allocation. The source in \cite{Chen} is degenerate: it outputs entangled photon pairs on the same wavelength. Thus, photons from an EPR pair cannot use the same fiber span in the same direction without routing ambiguity. This requires either time multiplexing (causing photon loss) or design of routing algorithms that accounts for this, along with path-dependent photon losses. Furthermore, although the source is broadband, when segmented into narrow-band channels, the rate of entangled photon pairs it generates per channel varies across the spectrum. Dynamic spectrum allocation techniques enable one such source to serve many consumer node pairs. However, their optimization must consider that channel assignments determine the EPR-pair rates received by the consumer node pairs. 

Fortunately, in our single-source setting, routing and spectrum allocation can be addressed separately. We present an algorithm to find the optimal routing paths under the constraints imposed by these network architectures. Our solution adapts Suurballe's algorithm \cite{Suurballe, Banerjee} to optimize the routes in polynomial time. Then, we build upon the classical approaches \cite{Simmons} to develop spectrum allocation strategies that ensure an equitable distribution of EPR-pair rates across the network, as in classical \cite{Chatterjee} and repeater-based quantum \cite{Li2024Survey,Chen2024Fairness,Yang2022Fairness,Zhao2023mixed} networks. Here we desire max-min fair spectrum allocation, where the minimum EPR-pair rate each node receives is maximized \cite{Saberi}. Unfortunately, as in classical optical networks \cite{Chatterjee}, this is an NP-hard integer linear program (ILP). Thus, we investigate the performance of various approximation algorithms, and compare them to the optimal ILP solution on a simple network. We identify two approximation algorithms that achieve close-to-optimal performance and analyze them on larger networks. Additionally, we report the Jain fairness index \cite{Jain} and the median EPR-pair rates. We find that a spectrum allocation approach that achieves higher minimum EPR-pair rate can perform significantly worse when these performance metrics as well as the computational resources are considered. We validate our approaches using numerical evaluation on a topology model based on an existing local exchange carrier (ILEC) network in Manhattan, New York, USA \cite{Yu, Li}, as well as larger synthetic topologies generated using the Watts-Strogatz model \cite{WattsStrogatz}. Additionally, we analyze the impact of source node placement for these topologies. We observe that the nodal degree has a significant impact on optimal EPR-pair source location.

Our routing and spectrum allocation protocols are a significant step towards implementing an intermediate-scale EPR-pair-distribution network on existing fiber deployments. This paper is organized as follows: we discuss prior work in the next section, while in Section \ref{sec:system_model} we overview and model the source and network architectures. We present our approaches for optimizing routing and spectrum allocation in Section \ref{sec:algorithims}, and compare them numerically in Section \ref{sec:results_and_discussion}. We discuss the implications of our results and future work in Section \ref{sec:conclusion}.

\section{Previous Work} \label{sec:previous_work}

Repeaterless entanglement distribution is viable in the near term, and has been experimentally demonstrated on small networks. For example, \cite{Wengerowsky} implements a single-hop network from source to four consumer nodes using passive fixed-grid wavelength-division multiplexing (WDM). As routing and consumer bandwidth allocations are static, the source node in \cite{Wengerowsky} must be reconfigured when a new consumer node is added.
This is addressed using wavelength-selective switches (WSSs) in \cite{Zhu}. Similarly, \cite{Wang} proposes a quantum re-configurable add-drop multiplexer (q-ROADM) to improve the efficiency. 

However, single-hop networks in \cite{Wengerowsky,Zhu,Wang} require direct connections between the source and consumer nodes. This not only imposes a potentially impractical constraint, but also reduces resiliency. 
The latter is addressed in \cite{alshowkan2024}, which demonstrates that entanglement distribution can be continued via an alternate two-hop path upon failure of the direct link.
Although the simplicity of the network topology in \cite{alshowkan2024} precluded investigation of routing protocols, it showed the promise of quantum network deployment on existing fiber infrastructure. We note that all repeaterless techniques are subject to a fundamental constraint known as the repeaterless bound, which is due to the exponential decay of fiber transmittance with length \cite{takeoka2014,Pirandola2017}. 

Research into larger-scale networks has focused on quantum-repeater networks that circumvent the repeaterless bound by placing repeater nodes along the optical path \cite{muralidharan2016optimal}. Extensive simulation studies for both bipartite and multipartite entanglement distribution in quantum-repeater networks are available \cite{pant2019routing,patil2022entanglement, wang2022pre, panigrahy2023scalable, kaur2023distribution, sutcliffe2023multi, van2023entanglement}. Although quantum repeaters have been demonstrated in the laboratory \cite{PhysRevLett.qunatumRepeater,LangenfeldQuantumRepeater}, we are not aware of operational repeater-based quantum networks. 

Despite extensive studies of quantum-repeater networks \cite{Li2024Survey}, fair quantum resource allocation has been rarely considered. However, the Jain fairness index \cite{Jain} has been applied in quantum-repeater networks \cite{Chen2024Fairness, Yang2022Fairness}, with \cite{Yang2022Fairness} also studying max-min fairness. In repeaterless networks, fairness has been largely unexplored, with \cite{Zhao2023mixed} beginning to address this gap by introducing optical switches into quantum-repeater networks, and considering fair path selection for these optically switched segments. Ultimately, the near-term feasibility and necessity of repeaterless networks demand a thorough exploration of fairness considerations, tackled in this work. 

\section{System Model}

\label{sec:system_model}

\begin{figure*}[!th]
\centering
  \subfloat[Network Optical Layout]{\includegraphics[width=0.9\textwidth]{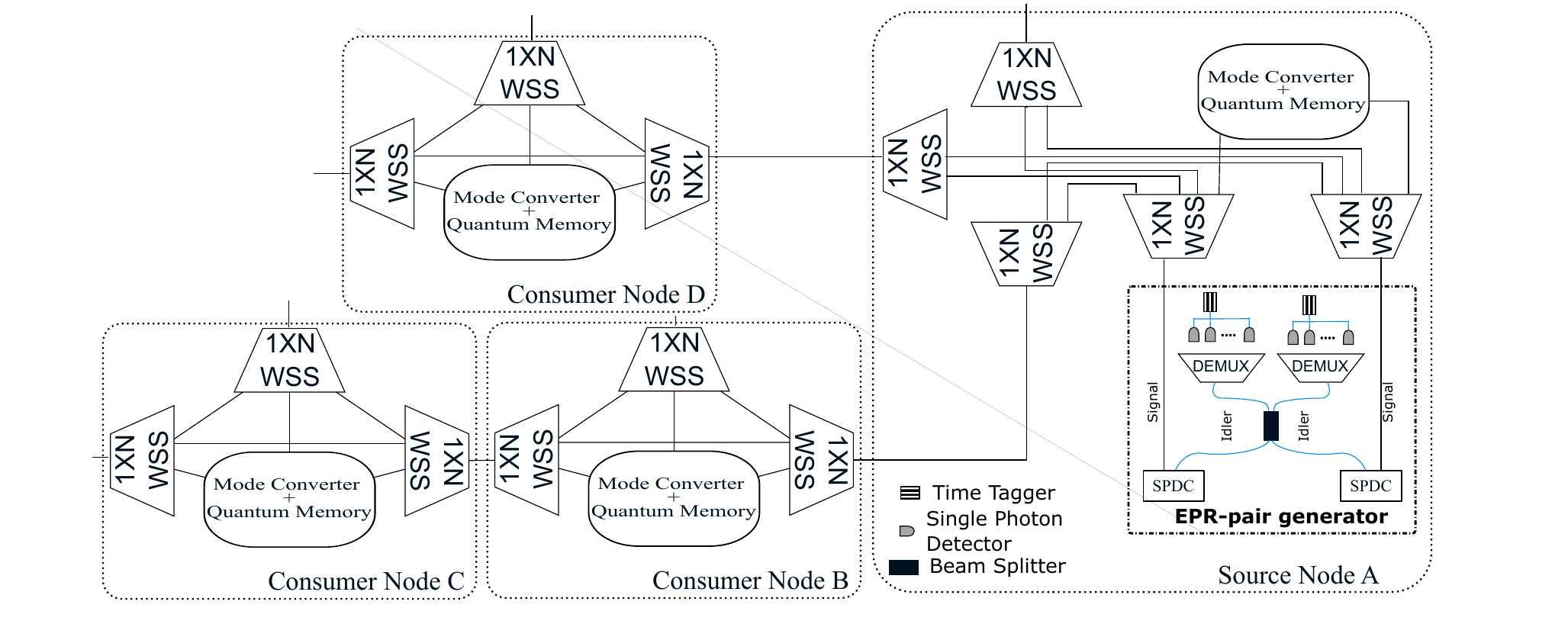} \label{fig:source_consumer_network_model}}
  
  \subfloat[Network Graph Model]{\includegraphics[width=0.9\textwidth]{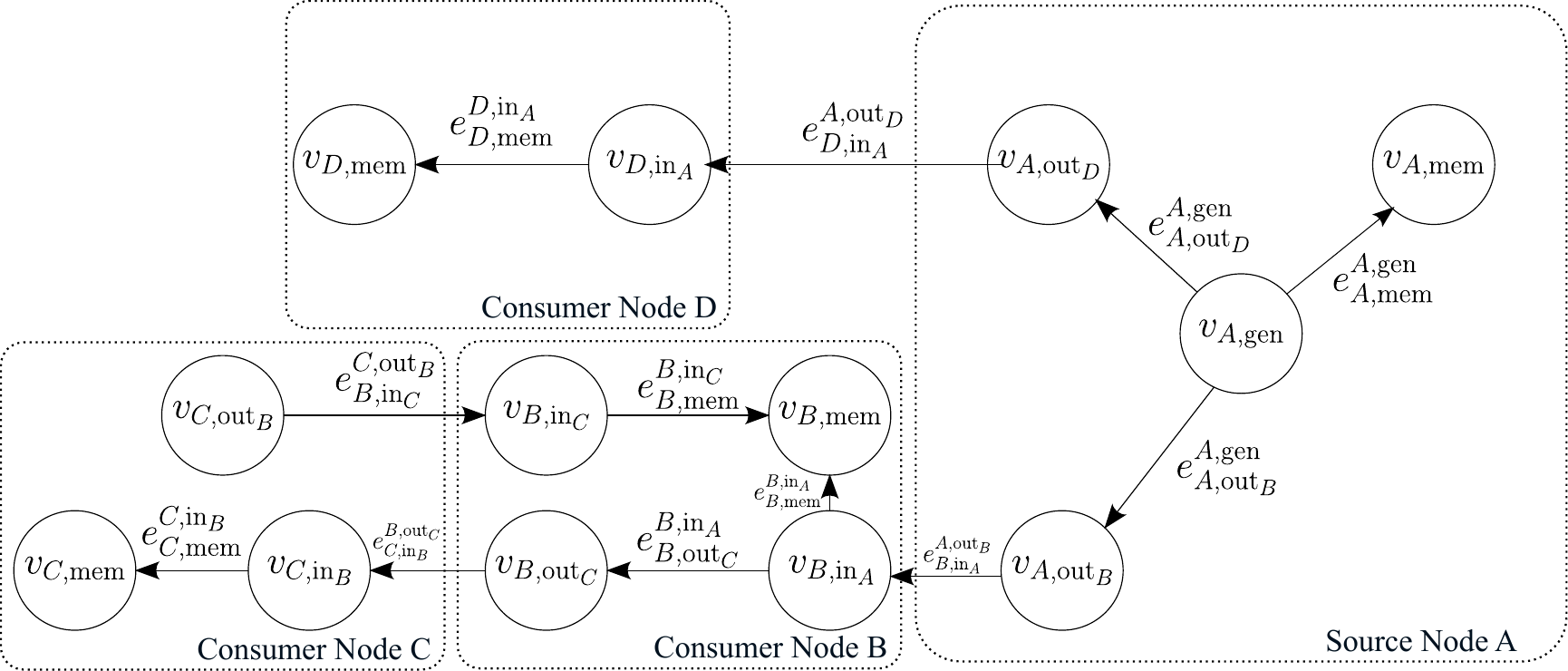} \label{fig:source_consumer_graph_model}}
  \caption{Correspondence between a network layout and its graph model. \protect\subref{fig:source_consumer_network_model} shows a network of source (A) and consumer nodes (B, C, and D). \protect\subref{fig:source_consumer_graph_model} shows the corresponding graph model.}
\label{fig:source_consumer}
\end{figure*}

\begin{table*}[t]
\centering
\begin{minipage}{\textwidth}  
\caption{Distance matrix (in km) for the map of Manhattan with ILEC nodes shown in Fig.~\ref{fig:manhattan_layout}.\label{table:manhattan_distance_matrix}}
\resizebox{\textwidth}{!}{  
\begin{tabular}{c c c c c c c c c c c c c c c c c c}  
\toprule
 & A & B & C & D & E & F & G & H & I & J & K & L & M & N & O & P & Q \\
\midrule
A & 0 & 0.304 & 1.184 & 2.032 & 3.744 & 5.2 & 4.352 & 5.776 & 6.096 & 5.84 & 7.232 & 7.04 & 8.8 & 9.12 & 10.688 & - & - \\
B & 0.304 & 0 & 0.912 & 1.712 & 3.488 & 5.056 & 4.048 & 5.488 & 5.936 & 5.296 & 6.848 & 6.656 & 8.496 & 8.816 & 10.32 & - & - \\
C & 1.184 & 0.912 & 0 & 2.336 & 2.08 & 3.328 & 2.304 & 3.728 & 4.192 & 3.904 & 5.296 & 5.04 & 6.752 & 7.216 & 9.664 & - & - \\
D & 2.032 & 1.712 & 2.336 & 0 & 2.224 & 3.36 & 2.368 & 3.728 & 2.192 & 4.0 & 5.392 & 4.992 & 6.848 & 7.216 & 8.768 & - & - \\
E & 3.744 & 3.488 & 2.08 & 2.224 & 0 & 1.44 & 1.6 & 2.448 & 2.624 & 1.968 & 3.472 & 3.728 & 5.28 & 5.312 & 6.88 & - & - \\
F & 5.2 & 5.056 & 3.328 & 3.36 & 1.44 & 0 & 1.696 & 1.536 & 1.36 & 0.544 & 2.0 & 2.528 & 4.0 & 3.872 & 5.456 & - & - \\
G & 4.352 & 4.048 & 2.304 & 2.368 & 1.6 & 1.696 & 0 & 1.408 & 1.888 & 2.112 & 3.312 & 2.624 & 4.496 & 5.056 & 6.496 & - & - \\
H & 5.776 & 5.488 & 3.728 & 3.728 & 2.448 & 1.536 & 1.408 & 0 & 0.624 & 1.408 & 2.176 & 1.28 & 3.04 & 3.696 & 5.152 & - & - \\
I & 6.096 & 5.936 & 4.192 & 2.192 & 2.624 & 1.36 & 1.888 & 0.624 & 0 & 1.12 & 1.552 & 1.264 & 2.704 & 3.12 & 4.576 & - & - \\
J & 5.84 & 5.296 & 3.904 & 4.0 & 1.968 & 0.544 & 2.112 & 1.408 & 1.12 & 0 & 1.376 & 2.288 & 3.424 & 3.296 & 4.832 & - & - \\
K & 7.232 & 6.848 & 5.296 & 5.392 & 3.472 & 2.0 & 3.312 & 2.176 & 1.552 & 1.376 & 0 & 2.208 & 2.56 & 1.92 & 3.472 & - & - \\
L & 7.04 & 6.656 & 5.04 & 4.992 & 3.728 & 2.528 & 2.624 & 1.28 & 1.264 & 2.288 & 2.208 & 0 & 1.872 & 3.2 & 4.256 & - & - \\
M & 8.8 & 8.496 & 6.752 & 6.848 & 5.28 & 4.0 & 4.496 & 3.04 & 2.704 & 3.424 & 2.56 & 1.872 & 0 & 4.8 & 6.368 & 2.96 & 6.096 \\
N & 9.12 & 8.816 & 7.216 & 7.216 & 5.312 & 3.872 & 5.056 & 3.696 & 3.12 & 3.296 & 1.92 & 3.2 & 4.8 & 0 & 1.536 & - & 5.856 \\
O & 10.688 & 10.32 & 9.664 & 8.768 & 6.88 & 5.456 & 6.496 & 5.152 & 4.576 & 4.832 & 3.472 & 4.256 & 6.368 & 1.536 & 0 & - & 4.368 \\
P & - & - & - & - & - & - & - & - & - & - & - & - & 2.96 & - & - & 0 & 3.04 \\
Q & - & - & - & - & - & - & - & - & - & - & - & - & 6.096 & 5.856 & 4.368 & 3.04 & 0 \\
\end{tabular}
}
\end{minipage}

\end{table*}

\begin{figure}[!t]
\centering
\includegraphics[width=\columnwidth]{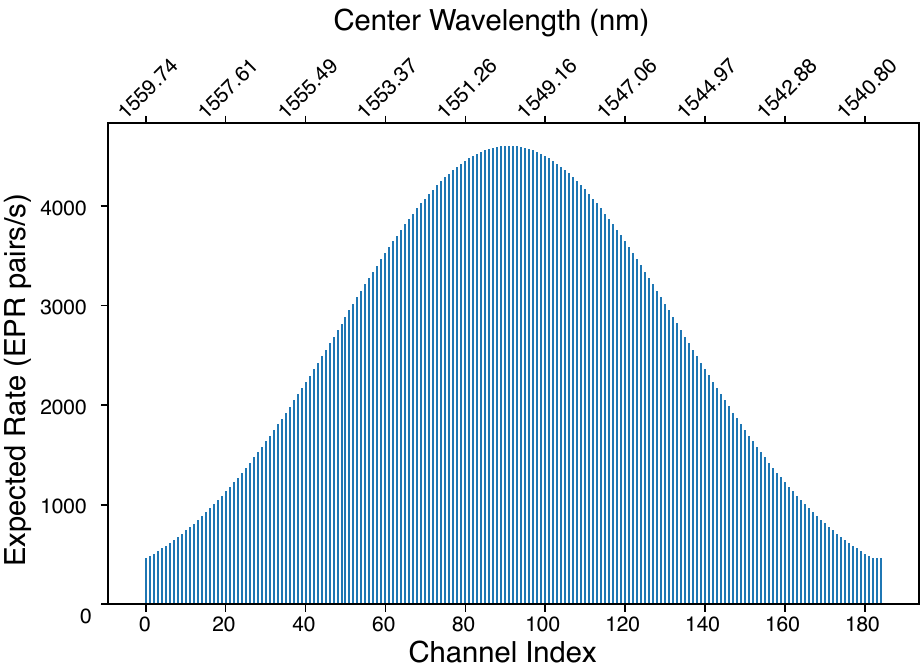}
\caption{Rate of EPR-pair generation in $m=185$ channels, each $B_{\text{c}}=11$ GHz wide, with center frequencies separated by $B_{\Delta}=13.135$ GHz, derived from \cite{shapiro2024entanglementsourcequantummemory}, used for simple and ILEC networks. The bottom-axis label shows the channel indices and the top-axis label shows the center wavelength of each channel. The highest and lowest per-channel rates are \num{4584} and \num{458} EPR pairs/s, respectively.}
\label{fig:dist_185}
\end{figure}

\subsection{Broadband Degenerate EPR-pair Generation}
\label{subsec:epr_gen}

The EPR-pair generation is depicted in Fig.~\ref{fig:source_consumer_network_model} as `EPR-pair generator' within `Source Node A.' We address Fig. \ref{fig:source_consumer_graph_model} in Section \ref{subsec:network_arch}. We assume the availability of a broadband, quasi-deterministic EPR-pair source. An example of such is the zero-added loss entangled-photon multiplexing (ZALM) scheme \cite{Chen}. It employs dual spontaneous parametric down-conversion (SPDC) processes, taking advantage of their broadband output. ZALM heralds polarization-entangled photon pairs in time and frequency via the following wavelength-demultiplexed Bell state measurement (BSM): the idler photons generated by both SPDC processes are interfered at a beamsplitter, wavelength-demultiplexed, and photodetected. The photon coincidence counts occurring at the same wavelength for two idler photons then herald entanglement of the signal photons. The corresponding heralded signal photons of now known and identical wavelength are directed through a WDM system with wavelength-selective add-drop capability. Detailed mathematical analysis of ZALM architecture is available in \cite{shapiro2024entanglementsourcequantummemory}.

We divide the $B_{\text{s}}=2.430$ THz spectrum segment centered at 1550 nm wavelength (1540.331--1559.794 nm or 192.333--194.763 THz) into $m$ channels, each $B_{\text{c}}$ Hz wide, with center frequencies separated by $B_{\Delta}$ Hz.
Fig. \ref{fig:dist_185} plots the expected EPR-pair rates generated by the source on each channel, for $m=185$ channels, $B_{\text{c}}=11$ GHz, and $B_{\Delta}=13.135$ GHz. The details of its calculation, which relies on the mathematical analysis provided in \cite{shapiro2024entanglementsourcequantummemory}, is in the Appendix. Channelization is further discussed in Section \ref{subsec:Channelization}.


Due to the Gaussian output spectrum from the SPDC sources, the EPR-pair generation rate per second for channels near 1550 nm is higher than for those on the edges of the spectrum. The output spectrum from this source can be routed and distributed across the network using WDM routing techniques like those that have been developed for classical optical networks \cite{Simmons}. Note that our analysis can be adapted to other methods of generating degenerate EPR pairs.

\subsection{Loss and Noise}
\label{subsec:loss_model}
We focus on the reduction in EPR-pair rates from losses in optical fiber links and WSSs because they dominate over thermal noise and other channel imperfections on links dedicated to quantum signal. WSS and fiber losses are characterized in Sections \ref{subsec:node_arch} and  \ref{subsec:network_arch}, respectively.

\subsection{Node Architecture}
\label{subsec:node_arch}

A source node supplies EPR pairs to consumer nodes in the network. The source node is also a consumer node: it has a quantum memory to store entanglement. However, consumer nodes cannot generate EPR pairs. Fig.~\ref{fig:source_consumer_network_model} diagrams the architecture of the source node and three identical consumer nodes, with the additional circuitry to generate EPR pairs shown in the `EPR-pair generator' within `Source Node A'. For all nodes, we assume identical quantum memory insertion loss that does not depend on the EPR pair center wavelength. Thus, it has no impact on routing and spectrum allocation. Furthermore, this loss is negligibly small relative to WSS and fiber losses.

Each photon of the generated EPR pair is directed by the source into a separate fiber. The node is built around $1\times N$ WSSs, whose role is to route wavebands towards different consumer nodes or else towards its own quantum memory bank. These wavebands group the source wavelength channels, e.g., those depicted in Fig.~\ref{fig:dist_185}. Information heralded by the EPR-pair generation process (including the channel and timestamp of the generated pair) is transmitted along a classical network that is not depicted here. A consumer node lacks EPR-pair generation capability but has all the other components of the source node. The block diagrams for both the source and consumer nodes are in Fig.~\ref {fig:source_consumer_network_model}. 

Measured insertion loss $l_{\text{WSS}}$ on Lumentum's TrueFlex Twin WSS ranges from 4 dB to 8 dB \cite{Lumentum}. Hence we analyze EPR-pair rate for two values of WSS loss: $l_{\text{WSS}}\in\{4,8\}$ dB. While they add significant loss, we note that WSSs are currently manufactured for use in classical networks and, thus, are not optimized for loss reduction. Other wavelength management and switching devices can achieve a loss of 2 dB \cite{Lumentum_lowloss}. Here we employ wavelength-independent loss $l$ in dB that is related to power transmittance by $\eta=10^{-l/10}$. 

\subsection{Network Topologies}
\label{subsec:net_topologies}
\begin{figure}[!t]
\centering
\includegraphics[width=\columnwidth]{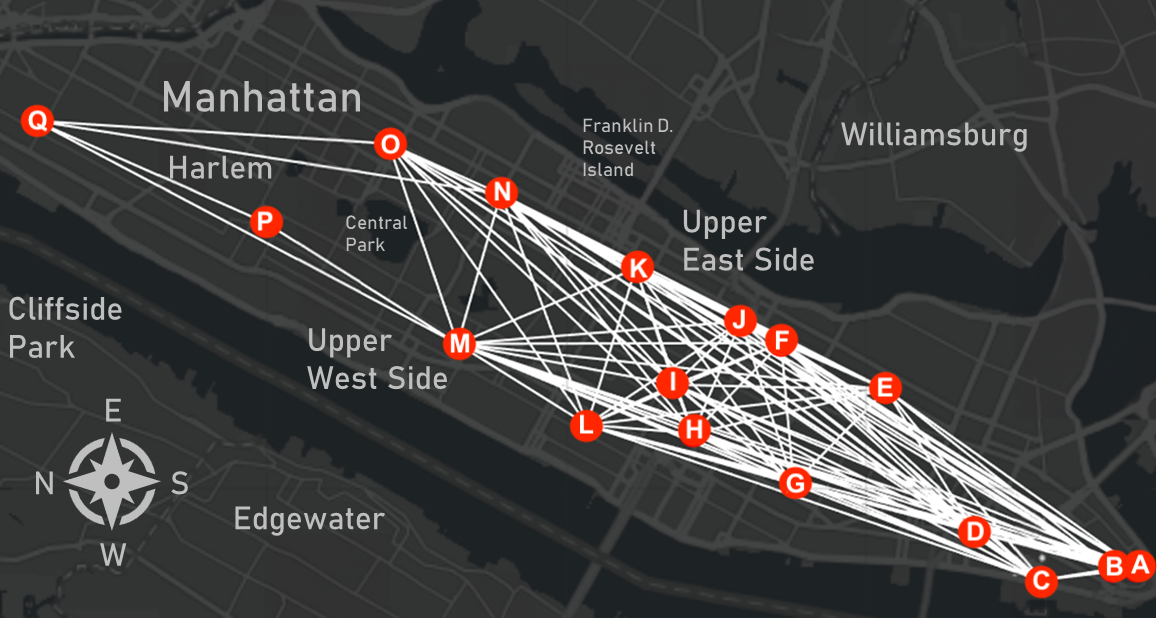}
\caption{A map of Manhattan with ILEC nodes and links overlaid. The distance matrix is given in Table \ref{table:manhattan_distance_matrix}.}
\label{fig:manhattan_layout}
\end{figure}

\begin{figure}[!t]
\centering
\includegraphics[width=.85\columnwidth]{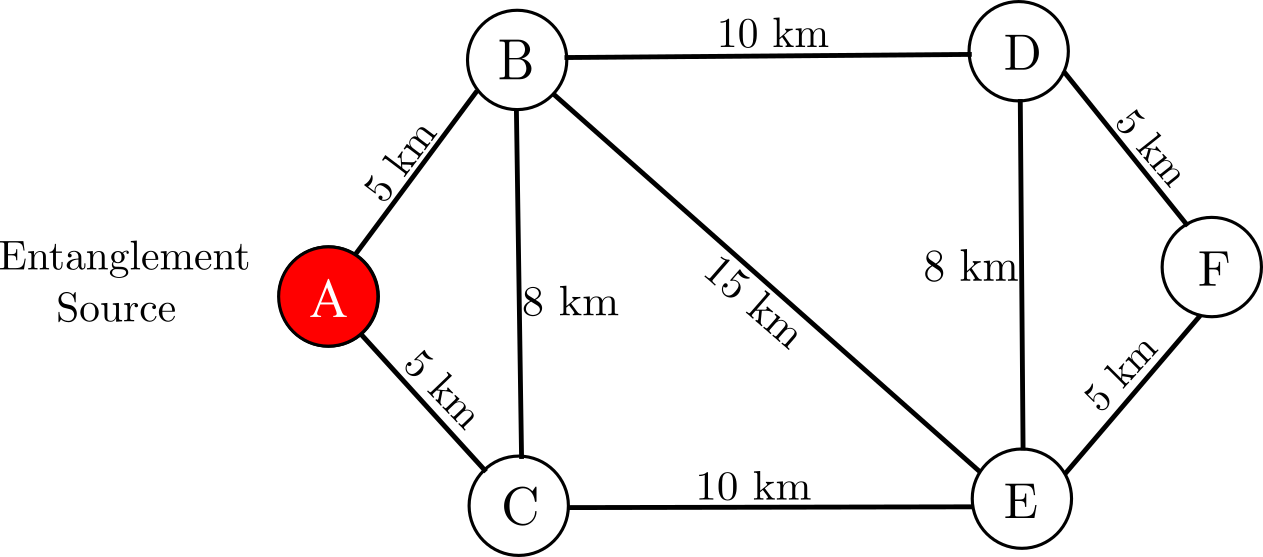}
\caption{Topology of the simple network. The EPR-pair source is at Node A.}
\label{fig:simple_layout}
\end{figure}

We analyze the topology model of an existing incumbent local exchange carrier (ILEC) node map of Manhattan \cite{Yu, Li}. This topology contains $n=17$ ILEC sites, with each site connected to between 2 and 16 other nodes. The layout of these nodes is shown in Fig.~\ref{fig:manhattan_layout}. The locations of ILEC sites are public but are not readily available.  Thus, we include an adjacency matrix with as-the-crow-flies distances in Table \ref{table:manhattan_distance_matrix}. While this is the reference topology for validating the performance of our approximation algorithms, the comparison with an optimal ILP solution is restricted to a smaller network topology with $n=6$ nodes, shown in Fig.~\ref{fig:simple_layout}, because the optimal fair allocation of the EPR-pair rate is NP-hard.

Additionally, we study the impact of network size on the performance by employing random Watts-Strogatz graphs \cite{WattsStrogatz}. These are generated as follows: first, $n$ nodes are generated and connected to their $k$ nearest neighbors symmetrically in a ring. Then, for every node, each edge connected to $k/2$ rightmost neighbors is rewired with probability $\beta$ to a different, randomly chosen node. These graphs are useful in network science due to their small-world properties. We consider the parameters $n \in \{10, 20, 30, 40\}$, $\beta \in \{0.2, 0.5, 0.8\}$, and $\frac{k}{n} \in  \bigl\{ \frac{1}{5}, \frac{2}{5}, \frac{3}{5}, \frac{4}{5} \bigl\} $.

\subsection{Traffic Model}
\label{subsec:traffic_model}

We consider a network with $n$ consumer nodes where one consumer node is also a source that generates EPR pairs. We defer the study of multi-source entanglement distribution networks to future work. We assume that all consumer node pairs in the network request EPR pairs from the source simultaneously. This results in $\kappa=\binom{n}{2}=n(n-1)/2$ node pairs and requests. We note, however, that, if only a subset of these $\kappa$ requests needs to be satisfied, the solutions presented subsequently apply unchanged.

\subsection{Channelization}
\label{subsec:Channelization}

For analysis of ILEC and simple topologies, we employ $m=185$ channels with rates of EPR-pair generation in Fig.~\ref{fig:dist_185}, whose calculation is described in the Appendix. For larger Watts-Strogatz networks $m=185$ channels are insufficient, because each of $\kappa=\binom{n}{2}=n(n-1)/2$ node pairs needs to be assigned at least one channel. We use $m = \lfloor 1.36\binom{n}{2} \rfloor$ since $185/\binom{17}{2} = 1.36$ channels per node pair are available in the ILEC network. This results in channel widths of 33.361 GHz when analyzing Watts-Strogatz graphs with $n = 10$ and 1.920 GHz with $n = 40$. We maintain a constant value of $\frac{1}{\kappa}\sum_{x=1}^m\bar{n}_{x}$ across differently-sized Watts-Strogatz topologies, where $\bar{n}_{x}$ is the EPR-pair rate generated in channel $x$.
While this may yield physically-impossible EPR-pair rates, it allows a fair performance comparison for networks with varying $n$.

\subsection{Network Architecture}
\label{subsec:network_arch}

The deployed fiber link lengths between nodes in the ILEC topology depicted in Fig.~\ref{fig:manhattan_layout} are unknown.  Thus, we use direct `as the crow flies' distances provided in Table \ref{table:manhattan_distance_matrix} as a proxy. The fiber link lengths between nodes of the simple network are shown in Fig.~\ref{fig:simple_layout}. In Watts-Strogatz networks we set all fiber link lengths to 5 km.

Standard single-mode fiber is assumed on each link. We employ a higher loss coefficient of $\alpha=0.4$ dB/km than typical fiber loss at 1550 nm (found in, e.g.,\cite{Kingfisher}) to account for higher losses and longer run lengths characteristic of metro fiber plant. Each wavelength channel is assigned to a single consumer node pair. The wavelength routing mechanism follows a circuit-switching approach. The routes serving different sets of node pairs do not interfere with one another, however, photons from a particular channel cannot be directed to both nodes of a pair via the same fiber in the same direction, as this results in a routing ambiguity. Therefore, we only consider networks that allow disjoint light-paths from the source to each of the $\kappa$ node pairs. Edge-disjoint paths can be absent only if the minimum cut of the graph is less than two. We eliminate such  instances of random Watts-Strogatz topologies.

\subsection{Network Graph Model}
\label{subsec:network_model}

\begin{table}[H]
\centering
\caption{Network Model Notation}
\renewcommand{\arraystretch}{1.5} 
\begin{tabular}{cm{0.35\textwidth}} 
\hline
\textbf{Notation} & \textbf{Description} \\ \hline
$v_{i,\text{out}_j}$ & Vertex for node $i$ that routes photons to node $j$. \\ 
$v_{i,\text{in}_j}$ & Vertex for node $i$ that receives photons from node $j$. \\
$v_{i,\text{mem}}$ & Vertex for node $i$ used to represent its quantum memory. \\ 
$v_{i,\text{gen}}$ & Vertex for node $i$ used to represent its EPR-pair generator (only present if $i$ is a source node) \\ 
$e^{x}_{y}$ & Directed edge from vertex $x$ to $y$. Note that vertex $x$ and $y$ have two internal indexes, e.g. $x=(i,\text{out}_j)$. \\ \hline
\end{tabular}
\label{table:notation}
\end{table}

We represent a network as a graph denoted by $\mathcal{G}=(\mathcal{V},\mathcal{E})$, where $\mathcal{V}$ and $\mathcal{E}$ are the sets of vertices and directed edges, respectively. Our notation is summarized in Table \ref{table:notation}. We note that $\mathcal{G}$ represents each network node with multiple vertices and edges. In Fig.~\ref{fig:source_consumer_graph_model}, we provide a `translation' to $\mathcal{G}$ of the optical network layout in Fig.~\ref{fig:source_consumer_network_model} as an example.  
We also define a map $w:\mathcal{E}\to\mathbb{R}$ that assigns photon losses (in dB) as edge weights. 
Mathematical construction of $\mathcal{G}$ for the network topologies described in Section \ref{subsec:net_topologies} is as follows:

\begin{itemize}
\item For each pair $(i,j)$ of connected consumer nodes we add the following directed edges and the corresponding vertices: $e^{i,\text{out}_j}_{j,\text{in}_i}\equiv\left(v_{i,\text{out}_j}, v_{j,\text{in}_i}\right)$ and $e^{j,\text{out}_i}_{i,\text{in}_j}\equiv\left(v_{j,\text{out}_i}, v_{i,\text{in}_j}\right)$ to $\mathcal{G}$. Hence each vertex is indexed by the node it belongs to, and by the role of that vertex. 
Vertices that serve input/output roles have the name of the corresponding external node as a subscript. The weight of the edges is $w\left(e^{i,\text{out}_j}_{j,\text{in}_i}\right)=w\left(e^{j,\text{out}_i}_{i,\text{in}_j}\right)=\alpha\times d(i,j)$, where $d(i,j)$ is the distance (in km) between nodes $i$ and $j$, and $\alpha$ is optical fiber loss (in dB/km) discussed in Section \ref{subsec:network_arch}.
\item For each consumer node $i$, we iterate over all nodes $j,k$ that connect to $i$, and add edges $e^{i,\text{in}_j}_{i,\text{out}_k}\equiv \left(v_{i,\text{in}_j}, v_{i,\text{out}_k}\right)$ to $\mathcal{E}$. This captures the consumer nodes' internal connections between incoming and outgoing ports. Note that this edge is skipped if $k$ is a source node, as the source node has no incoming edges. Since the photons routed through a consumer node must traverse two WSSes, the weight of these edges is $w\left(e^{i,\text{in}_j}_{i,\text{out}_k}\right)=2l_{\rm WSS}$, as discussed in Section \ref{subsec:network_arch}.  Furthermore, we also add edges $e^{i,\text{in}_j}_{i,\text{mem}}\equiv \left(v_{i,\text{in}_j}, v_{i, \text{mem}}\right)$ describing internal connections to node $i$'s quantum memory to $\mathcal{E}$, and the corresponding vertices to $\mathcal{V}$. Since only one WSS is traversed in this case, $w\left(e^{i,\text{in}_j}_{i,\text{mem}}\right)=l_{\rm WSS}$. 
\item For the source node $s$, we iterate over all nodes $j$ that connect to $s$, and add edges $e^{s,\text{out}_j}_{j,\text{in}_s}\equiv \left(v_{s,\text{out}_j}, v_{j, \text{in}_s}\right)$ to $\mathcal{E}$ and corresponding vertices to $\mathcal{V}$. The weight of these edges is $w\left(e^{s,\text{out}_j}_{j,\text{in}_s}\right)=\alpha\times d(s,j)$. Consumer node's incoming vertices $v_{j,\text{in}_s}$ are connected to outgoing vertices and quantum memories as described above. Finally, we add edges $e^{s,\text{gen}}_{s,\text{out}_j}\equiv \left(v_{s,\text{gen}}, v_{s,\text{out}_j}\right)$ and $e^{s,\text{gen}}_{s,\text{mem}}\equiv \left(v_{s,\text{gen}}, v_{s,\text{mem}}\right)$ from vertex $v_{s,\text{gen}}$ describing EPR-pair source to all outgoing ports and vertex $v_{s, \text{mem}}$ describing source's own quantum memory. These edges' weights are $w\left(e^{s,\text{gen}}_{s,\text{out}_j}\right)=2l_{\rm WSS}$ and $w\left(e^{s,\text{gen}}_{s,\text{mem}}\right)=l_{\rm WSS}$, per above. Note that the source node has no incoming ports. 

\end{itemize}
The total loss on a path from source to a consumer node $i$ is the sum of weights of the edges connecting $v_{s, \text{gen}}$ to $v_{i,\text{mem}}$.

\subsection{Max-min Fair Allocation of Received EPR-pair Rates}
\label{subsec:egal_fairness}
We seek max-min, or egalitarian, fairness, by maximizing the minimum rate of EPR pairs received by all $\kappa = n(n-1)/2$ pairs $(i,j)$ of $n$ nodes \cite{Saberi}. Let $l_{(i,j)}$ be the total loss (in dB) from the source to nodes $(i,j)$. That is, $l_{(i,j)}$ is the sum of losses on the disjoint paths from source to nodes $i$ and $j$, per Section \ref{subsec:network_model}. Then, transmittance $\eta_{(i,j)}=10^{-l_{(i,j)}/10}$ is the fraction of the entangled photon pairs that are received by $(i,j)$. Let $\mathcal{A}_{(i,j)}$ be the set of channels assigned to node pair $(i,j)$. Since each channel cannot be assigned to more than one node pair, the set $\mathcal{P}=\left\{\mathcal{A}_{(i,j)}:i,j=1,\ldots,n, i \neq j \right\}$ partitions the $m$ available channels. Let $\bar{n}_{x}$ be the EPR-pair rate generated in channel $x$. The EPR-pair rate received by node pair $(i,j)$ is then $\bar{n}_{(i,j)}=\eta_{(i, j)}\sum_{x\in \mathcal{A}_{(i,j)}} \bar{n}_{x}$ and the max-min fair allocation involves the following optimization: 
   $\max_{\mathcal{P}} \min_{(i, j)} \bar{n}_{(i,j)}$.

\section{Algorithms}
\label{sec:algorithims}
Orthogonality of sets $\mathcal{A}_{(i,j)}$ allows treating routing and spectrum allocation problems separately, as discussed next. 

\subsection{Optimal routing}
\label{sec:paths}

Unlike standard networks, our source-in-the-middle entanglement distribution system described in Section \ref{sec:system_model} requires two disjoint light paths from source $s$ to nodes $i$ and $j$ that minimize total loss $l_{(i,j)}$ for each pair $(i,j)$ in the network. Per Section \ref{subsec:network_model}, this translates to finding edge-disjoint routes in $\mathcal{G}$ from $v_{s,\text{gen}}$ to $v_{i, \text{mem}}$ and  $v_{j,\text{mem}}$ minimizing the sum of weights of these paths. 
To this end, we use Suurballe's algorithm \cite{Suurballe, Banerjee} as follows: for each consumer pair $(i,j)$ we add a dummy vertex $v_{(i,j),\text{d}}$ to $\mathcal{V}$ and dummy zero-weighted edges: $e^{i,\text{mem}}_{(i,j),\text{d}}\equiv \left(v_{i,\text{mem}}, v_{(i,j),\text{d}}\right)$ and $e^{j,\text{mem}}_{(i,j),\text{d}}\equiv \left(v_{j,\text{mem}}, v_{(i,j),\text{d}}\right)$ to $\mathcal{E}$. Suurballe's algorithm yields two edge-disjoint paths of minimum total weight between $v_{s, \text{gen}}$ and $v_{(i,j),\text{d}}$.
Removing dummy vertices and edges returns edge-disjoint paths of minimum total weight from $v_{s, \text{gen}}$ to $v_{i, \text{mem}}$ and $v_{j,\text{mem}}$ for all pairs $(i,j)$.
Suurballe's algorithm's run-time is polynomial in graph size.

\subsection{Spectrum Allocation Strategies}
\label{sec:allocation_strategies}
Let $X$ be an $m\times \kappa$ binary matrix with $X_{x,(i,j)}=1$ if channel $x$ is assigned to node pair $(i,j)$ and zero otherwise (note that the pair $(i,j)$ indexes columns of $X$). Formally, $X_{x,(i,j)}=\left\{1~\text{if}~x\in\mathcal{A}_{(i,j)}; 0~\text{else}\right\}$.
Also define an $\kappa\times \kappa$ diagonal matrix $\Lambda$ with transmittances $\eta_{(i,j)}^\ast$ of optimal routes (see Section \ref{sec:paths}) from source to each node pair $(i,j)$ on the diagonal and a vector $N = \left[\bar{n}_{1}, \ldots, \bar{n}_{m}\right]$ of EPR-pair-generation rates in each channel (see Section \ref{subsec:egal_fairness}).
For some $X$, the rate of EPR pairs received by $(i,j)$ is $\bar{n}_{(i,j)}=\left[NX\Lambda\right]_{(i,j)}$, the $(i,j)^{\text{th}}$ entry of vector $NX\Lambda$.
Finding an optimal spectrum allocation matrix $X$ is a well-known problem in optical networking \cite{Chatterjee}. Here we focus on maintaining max-min fairness in source-in-the-middle entanglement distribution.

\subsubsection{Optimal Assignment}

The following integer linear program (ILP) yields the optimal max-min fair solution:
\begin{IEEEeqnarray}{ll}
\label{eq:ilp}
\IEEEyesnumber  \IEEEyessubnumber*
\max_X T \text{~s.t.~}&\sum_{\myatop{i,j=1}{i\neq j}}^{n} X_{x,(i,j)} = 1, \forall x=1,\ldots,m \label{eq:ilp_second}\\
&\left[NX\Lambda\right]_{(i,j)} \geq T, \forall i,j=1,\ldots,n,i\neq j,  \label{eq:ilp_third}
\end{IEEEeqnarray}
where constraint \eqref{eq:ilp_second} enforces that each channel is assigned only once and \eqref{eq:ilp_third} ensures that each node pair receives EPR-pair rate of at least $T$. 

The routing scheme in our scenario implicitly enforces wavelength contiguity constraints, as wavelengths cannot be switched at intermediate nodes.
This contrasts with classical optical networks, where optimal spectrum allocation has to explicitly enforce them.
Additionally, unlike classical networks that allow fractional channel allocation, source-in-the-middle entanglement distribution requires discrete channel assignment to entangle two particular quantum memories.  
This necessitates solving an NP-hard ILP problem. 
Hence, we consider approximations.

\subsubsection{Round Robin \texorpdfstring{\cite{Aziz}}{}}
Consumer node pairs, sorted in descending order of loss, are assigned channels in a cyclic order. The available channel with the highest EPR-pair rate is selected at each step. Algorithm~\ref{alg:roundrobin} provides the pseudocode. The time complexity of this algorithm is determined by sorting the $\kappa$ node pairs and $m$ channels. Since $m\geq\kappa$, the overall time complexity is $\mathcal{O}(m\log{m})$.

\begin{algorithm}
\small
\begin{algorithmic}[1]
\phase{Main Procedure}
\end{algorithmic}
\begin{algorithmic}[1]
\Procedure{RoundRobin}{$\Lambda, N$}
\LineComment{Inputs and outputs are specified in Section
\ref{sec:allocation_strategies}}
\State \textbf{Input:} 
\State \hspace{\algorithmicindent} $\Lambda$: $\kappa\times\kappa$ diagonal matrix of transmittances from source to consumer node pairs
\State \hspace{\algorithmicindent} $N$: $m\times 1$ row vector with channel EPR-pair rates
\State \textbf{Output:} 
\State \hspace{\algorithmicindent} $X$: $m\times\kappa$ binary matrix with $X_{x,(i,j)}=1$ if channel $x$ is assigned to node pair $(i,j)$
\State $X \gets \Call{zeros}{m, \kappa}$
\State $\eta \gets \Lambda.diagonal$ \Comment{Vector of transmittances}
\State $\eta_{\text{sorted}}, I_{\eta} \gets \Call{SortAscending}{\eta}$ 
\State $N_{\text{sorted}}, I_{N} \gets \Call{SortDescending}{N}$ 

\State $i \gets 0$ \Comment{Tracks channel index at current iteration}
\While{$i < m$}
    \State $j \gets i \bmod \kappa$ \Comment{next lowest transmittance picks the next highest channel}
    \State $X[I_{N}[i], I_{\eta}[j]] \gets 1$
    \State $i \gets i + 1$
\EndWhile
\State \Return $X$
\EndProcedure
\end{algorithmic}
\begin{algorithmic}[1]
\phase{Auxiliary procedures}
\end{algorithmic}
\begin{algorithmic}[1]
\Procedure{SortAscending}{$a_{\text{unsorted}}$}
    \State \textbf{Input:} $a_{\text{unsorted}}$: Unsorted array of comparable items
    \State \textbf{Output:}
    \State \hspace{\algorithmicindent} $a_{\text{sorted}}$: Sorted array in ascending order 
    \State \hspace{\algorithmicindent} $I$: Indices mapping the sorted array to the original
\EndProcedure
\vspace{1em} 
\Procedure{SortDescending}{$a_{\text{unsorted}}$}
    \State \textbf{Input:} $a_{\text{unsorted}}$: Unsorted array of comparable items
    \State \textbf{Output:}
    \State \hspace{\algorithmicindent} $a_{\text{sorted}}$: Sorted array in descending order
    \State \hspace{\algorithmicindent} $I$: Indices mapping the sorted array to the original
\EndProcedure
\end{algorithmic}
\caption{Round Robin}\label{alg:roundrobin}
\end{algorithm}

\subsubsection{First Fit \texorpdfstring{\cite{Chatterjee}}{}}
We assign channels sequentially to a node pair. If EPR-pair rate $T$ is reached, then we repeat for the next node pair. The ordering of consumer node pairs follows a descending order based on loss. We restart with a smaller $T$ if channels are exhausted before all node pairs attain EPR-pair rate $T$. The maximum value $T^\ast$ satisfied by this algorithm can be found via binary search. Algorithm~\ref{alg:firstfit} provides the pseudocode. Its time complexity is $\mathcal{O}\left(m\log\left({\frac{1}{\kappa}\sum_{x=1}^m \bar{n}_x}\right)\right)$.

\begin{algorithm*}
\small
\begin{algorithmic}[1]
\phase{Main Procedure}
\end{algorithmic}
\begin{algorithmic}[1]
    \Procedure{FirstFit}{$\Lambda, N$}
        \State \textbf{Input} and \textbf{Output} as defined in Algorithm \ref{alg:roundrobin}
        \State $\eta \gets \Lambda.diagonal$ \Comment{Vector of transmittances}
        \State $\eta_{\text{sorted}}, I_{\eta} \gets \Call{SortAscending}{\eta}$ \Comment{Defined in auxiliary procedures of Algorithm \ref{alg:roundrobin}}
        
        \State $T^\ast \gets \Call{BinarySearch}{low=0, high= \lceil{\frac{sum(N)}{\kappa}}\rceil, predicate=\textproc{FirstFitPredicate}}$

        \State $X, \_ \gets \Call{FirstFitAssignment}{T^\ast}$ \Comment{The second return value is ignored}
        \State \Return $X$
    \EndProcedure
\end{algorithmic}
\begin{algorithmic}[1]
\phase{Auxiliary procedures inheriting context from the main procedure}
\end{algorithmic}
\begin{multicols}{2}
\raggedcolumns
\begin{algorithmic}[1]
    \Procedure{FirstFitPredicate}{$T$}
        \State \textbf{Input:} $T$: Threshold value to compare against the assigned EPR-pair rates.
    \State \textbf{Output:} A boolean value indicating whether all node pairs meet the threshold $T$.

        \State $\_, A \gets \Call{FirstFitAssignment}{T}$ \Comment{The first return value is ignored}
        \For{$j \gets 0, \dots, (\kappa-1)$}
            \If{$A[j] < T$}
                \State \Return false
            \EndIf
        \EndFor
        \State \Return true
    \EndProcedure
    \vspace{1em} 
    \Procedure{FirstFitAssignment}{$T$}
        \State \textbf{Input:} $T$: Threshold value for the assignment of EPR-pair rates to node pairs.
    \State \textbf{Output:} 
    \State \hspace{\algorithmicindent} $X$: Binary matrix representing the assignments, where $X[i, j] = 1$ if channel $i$ is assigned to node pair $j$
    \State \hspace{\algorithmicindent} $A$: Array of assigned EPR-pair rates for each node pair, with $A[j]$ being the total rate for node pair $j$.

        \State $X \gets \Call{zeros}{m, \kappa}$
        \State $A \gets \Call{zeros}{1, \kappa}$ \Comment{Assigned EPR-pair rates for each node pair}
        \State $i \gets 0$ \Comment{Tracks channel index}
        \State $j \gets 0$ \Comment{Tracks node pair index}
        \While{$i < m \And j < \kappa$}
            \State $\hat{j} \gets I_{\eta}[j]$ \Comment{Node pair's index in unsorted list}
            \State $X[i, \hat{j}] \gets 1$
            \State $i \gets i + 1$
            \State $A[\hat{j}] \gets A[\hat{j}] + \eta_{\text{sorted}}[j]N[i]$
            \If{$A[\hat{j}] \geq T$}
                \State $j \gets j + 1$
            \EndIf
        \EndWhile
        \State \Return $X, A$
    \EndProcedure
\vspace{1em} 
\Procedure{BinarySearch}{$low, high, predicate$}
    \State \textbf{Input:} 
    \State \hspace{\algorithmicindent} $low$: Lower bound of the search range
    \State \hspace{\algorithmicindent} $high$: Upper bound of the search range
    \State \hspace{\algorithmicindent} $predicate$: Monotonic boolean function
    \State \textbf{Output:} The largest value $i$ between $low$ and $high$ such that $predicate(i)$ is true.
    \State $i \gets low$
    \While{$i < high$}
        \State $mid \gets \left\lfloor \frac{i + high}{2} \right\rfloor$
        \If{$predicate(mid)$}
            \State $i \gets mid + 1$
        \Else
            \State $high \gets mid$
        \EndIf
    \EndWhile
    \State \Return $i$
\EndProcedure
\end{algorithmic}
\end{multicols}
\caption{First Fit}\label{alg:firstfit}
\end{algorithm*}

\subsubsection{Modified Longest Processing Time First (LPT) \texorpdfstring{\cite{Graham, Deuermeyer}}{}}

We modify a well-known machine scheduling algorithm to greedily optimize for the max-min rather than min-max goal, akin to \cite{Wu}: each channel is assigned to a node pair which maximizes the current minimum received EPR-pair rate across the node pairs. While our experiments indicate that this approach performs well, we have not derived any analytical performance guarantees. The pseudocode for this algorithm is given in Algorithm~\ref{alg:mod_lpt}. Its time complexity is determined by sorting the $\kappa$ node pairs and $m$ channels, and finding the minimum value of $\kappa$ node pairs for $m-\kappa$ iterations. The time required for sorting the node pairs can be disregarded, as in the analysis of the Round Robin algorithm. Hence the overall time complexity is given as $\mathcal{O}(m\kappa + m\log m)$.

\begin{algorithm}
\small
\begin{algorithmic}[1]
\Procedure{ModifiedLPT}{$\Lambda, N$}
    \State \textbf{Input} and \textbf{Output} as defined in Algorithm \ref{alg:roundrobin}
    \State $X \gets \Call{zeros}{m, \kappa}$
    \State $\eta \gets \Lambda.diagonal$ \Comment{Vector of transmittances}
    \State $\eta_{\text{sorted}}, I_{\eta} \gets \Call{SortAscending}{\eta}$ \Comment{Defined in auxiliary procedures of Algorithm \ref{alg:roundrobin}}
    \State $N_{\text{sorted}}, I_{N} \gets \Call{SortDescending}{N}$ \Comment{Defined in auxiliary procedures of Algorithm \ref{alg:roundrobin}}
    \State $A \gets \Call{zeros}{1, \kappa}$ \Comment{Initial vector of assigned EPR-pair rates for each node pair}
    \LineComment{Assign a single channel to each node pair}
    \For{$j \gets 0, \dots, (\kappa-1)$}
        \State $X[I_{N}[j],I_{\eta}[j]] \gets 1$
        \State $A[I_{\eta}[j]] \gets A[I_{\eta}[j]] + \eta_{\text{sorted}}[j]N_{\text{sorted}}[j]$
    \EndFor
    \LineComment{Now assign each remaining channel greedily}
    \State $i \gets \kappa$ 
    \While{$i < m$}
        \State $j \gets \Call{ArgMin}{A}$ \Comment{Index of node-pair with lowest EPR-pair rate currently assigned}
        \State $X[I_{N}[i],j] \gets 1$
        \State $A[j] \gets A[j] + \eta[j]N_{\text{sorted}}[i]$
        \State $i \gets i + 1$
    \EndWhile
    \State \Return $X$
\EndProcedure
\end{algorithmic}
\caption{Modified LPT}\label{alg:mod_lpt}
\end{algorithm}

\subsubsection{Bez\'{a}kov\'{a} and Dani's \texorpdfstring{$1/(m-\kappa+1)$}{1/(m-k+1)}-approximation (BD) \texorpdfstring{\cite{Bezakova}}{}}
 This iterative polynomial-time algorithm converges to a solution that is guaranteed to be within $1/(m-\kappa+1)$ of the optimal max-min EPR-pair rate. We make two modifications: 1) instead of always assigning one channel to each node pair in each round, we allow skipping a channel assignment; 2) in each round, we prefer the assignment which minimizes the total assigned rate of EPR-pair generation. These are invoked as long as they do not impact the overall max-min EPR-pair rate, hence they can only increase the minimum received EPR-pair rate for all node pairs, all the while preserving the original approximation guarantee. The pseudocode for this algorithm is in Algorithm~\ref{alg:maxmin1}. It assigns each node pair one channel in the first round, and zero or one channels in subsequent rounds, and thus has to run for $m-\kappa + 1$ rounds in the worst case. In each round, we run a binary search taking a maximum of $\log(\max(N))$ iterations. For each iteration, the maximum matching can be found in $\mathcal{O}(m\kappa(m+\kappa) + (m + \kappa)^2)$ time using the Ford-Fulkerson algorithm \cite{Tardos}, which simplifies here to $\mathcal{O}(m^2\kappa)$ as $m \geq \kappa$. Finally, finding the minimum-weight matching takes $\mathcal{O}(\kappa\log(\kappa) + m\kappa)$ time, which simplifies to $\mathcal{O}(m\kappa)$. Thus, the total time complexity is $\mathcal{O}(g(m,\kappa))$, where
\begin{align}
    g(m,\kappa)&=(\log(\max(N))m^2\kappa + m\kappa)\times(m-\kappa + 1),
\end{align}
which simplifies to $\mathcal{O}(m^3\kappa\log(\max(N)))$.

\begin{algorithm*}
\small
\begin{algorithmic}[1]
\phase{Main Procedure}
\end{algorithmic}

\begin{algorithmic}[1]

\Procedure{BD}{$\Lambda, N$}
    \State \textbf{Input} and \textbf{Output} as defined in Algorithm \ref{alg:roundrobin}
    \State $\eta \gets \Lambda.diagonal$ \Comment{Vector of transmittances}
    \State $V_{\text{c}}, V_{\text{np}}, E \gets \Call{BipartiteGraph}{\eta, N}$
    \State $X \gets \Call{zeros}{m, \kappa}$,  $T^\ast \gets 0$, $A \gets \Call{zeros}{1, \kappa}$ \Comment{Vector $A$ tracks assigned EPR-pair rates for each node pair}
    \While{$\text{length}(V_{\text{c}}) \geq \text{length}(V_{\text{np}})$}
        \LineComment{\textproc{BinarySearch} is Algorithm \ref{alg:firstfit}'s aux.~procedure. All but the last argument to \textproc{MatchingExists} are predefined}
        \State $T^\ast \gets \Call{BinarySearch}{low=T^\ast, high= T^\ast + \lceil{\frac{\text{sum}(N)}{\kappa}}\rceil, predicate=\textproc{MatchingExists}((V_{\text{c}}, V_{\text{np}}, E), A, \_)}$ 
        \State $V_{\text{c}}^{\text{mod}}, V_{\text{np}}^{\text{mod}}, E^{\text{mod}} \gets \Call{ModifiedGraph}{(V_{\text{c}}, V_{\text{np}}, E), A, T^\ast}$
        \State $V_{\text{c}}^{\text{min}}, V_{\text{np}}^{\text{min}}, E^{\text{min}} \gets$ \Call{MinWeightMatching}{$(V_{\text{c}}^{\text{mod}}, V_{\text{np}}^{\text{mod}}, E^{\text{mod}})$} \Comment{Note: the original algorithm in \cite{Bezakova} uses arbitrary matching}
        
        \ForAll{$e$ in $E^{\text{min}}$}
            \State $X[i_{\text{c}}, i_{\text{np}}] \gets 1$ \Comment{$i_{\text{c}}$ is the index of $e.v_{\text{c}}$ in $N$, and $i_{\text{np}}$ is the index of $e.v_{\text{np}}$ in $\eta$}
            \State $V_{\text{c}} \gets V_{\text{c}} \setminus \{e.v_{\text{c}}\}$
            \State $E \gets E \setminus \{e\}$
            \State $A[i_{e.v_{\text{np}}}] \gets A[i_{e.v_{\text{np}}}] + e.w$ \Comment{$i_{e.v_{\text{np}}}$ is the index of $e.v_{\text{np}}$ in $\eta$}
        \EndFor
    \EndWhile
    \LineComment{The original algorithm in \cite{Bezakova} leaves the remaining channels unassigned. Here we assign them using \textproc{RoundRobin}.}
    \LineComment{Define $N_{\text{rem}}$ as the channel rates of the unassigned channels. Also define $I_{\text{rem}}$ as the index of these channels in $N$.}
    \State $X_{\text{RR}} \gets \Call{RoundRobin}{N_{\text{rem}},\Lambda}$
    \State $X[I_{\text{rem}}[j], :] \gets X_{\text{RR}}[j, :],\quad \forall j \in length(N_{\text{rem}})$
    \State \Return $X$
\EndProcedure
\end{algorithmic}
\begin{algorithmic}[1]
\phase{Auxiliary procedures inheriting context from the main procedure}
\end{algorithmic}

\begin{multicols}{2}
\raggedcolumns
\columnbreak
\begin{algorithmic}[1]
\Procedure{BipartiteGraph}{$\eta, N$}
    \State \textbf{Input:} 
    \State \hspace{\algorithmicindent} $\eta$: $\kappa \times 1$ vector of transmittances for each node-pair
    \State \hspace{\algorithmicindent} $N$: $m\times1$ vector of EPR-rates for each channel 
    \State \textbf{Output:} 
    \State \hspace{\algorithmicindent}  $V_{\text{c}}, V_{\text{np}}$: Vertex partitions corresponding to channels and node pairs
    \State \hspace{\algorithmicindent} $E$: edge set, where, for each $e\in E$, $e.v_{\text{c}}\in V_{\text{c}}$, $e.v_{\text{np}}\in V_{\text{np}}$, and weight $e.w$ the product of source EPR-pair rate of $e.v_{\text{c}}$ and transmittance to node pair in $e.v_{\text{np}}$
    \State $V_{\text{c}}, V_{\text{np}}, E \gets \Call{CompleteBipartiteGraph}{m, \kappa}$
    \For{$e$ in $E$}
        \State $e.w \gets \eta[i_{\text{c}}] \times N[i_{\text{np}}]$ \Comment{$i_{\text{c}}$ is the index of $e.v_{\text{c}}$ in $N$, and $i_{\text{np}}$ is the index of $e.v_{\text{np}}$ in $\eta$}
    \EndFor
    \State \Return $(V_{\text{c}}, V_{\text{np}}, E)$
\EndProcedure
\vspace{0.5em} 
\Procedure{MatchingExists}{$(V_{\text{c}}, V_{\text{np}}, E), A, T$}
    \State \textbf{Input:} 
    \State \hspace{\algorithmicindent} $(V_{\text{c}}, V_{\text{np}}, E)$: Bipartite graph's vertices and edges
    \State \hspace{\algorithmicindent} $A$: Assigned EPR-pair rates for each node pair
    \State \hspace{\algorithmicindent} $T$: Threshold
    \State \textbf{Output:} Boolean 
    \State \Return true if perfect matching exists in \Call{ModifiedGraph}{$(V_{\text{c}}, V_{\text{np}}, E), A, T$} \Comment{Checked with any max-cardinality algorithm, e.g., Ford-Fulkerson \cite{Tardos}}    
\EndProcedure
\vspace{0.5em} 
\Procedure{ModifiedGraph}{$(V_{\text{c}}, V_{\text{np}}, E), A, T$}
\LineComment{Removes node-pair vertices meeting $T$, updates $e.w$ to final EPR rates received by $e.v_{np}$ if assigned $e.v_{c}$, and removes edges that cannot satisfy demand.}
    \State \textbf{Input:} as defined for \textproc{MatchingExists}
    \State \textbf{Output:} $(V_{\text{c}}^{\text{mod}}, V_{\text{np}}^{\text{mod}}, E^{\text{mod}})$: Vertex subsets and edges of the modified bipartite graph. 
    \State $V_{\text{np}}^{\text{mod}} \gets \{v \in V_{\text{np}} \mid A[i_{\text{np}}] < T\}$ \Comment{$i_{\text{np}}$ is the index of $e.v_{\text{np}}$ in $\eta$}
    \State $E^{\text{mod}} \gets \{e \mid e \in E \text{~with~} e.w \gets e.w + A[i_{e.v_{\text{np}}}]\}$
    \State $E^{\text{mod}} \gets \{e \in E^{\text{mod}} \mid e.w \geq T\}$
    \State $V_{\text{c}}^{\text{mod}} \gets \{v \in V_{\text{c}} \mid \exists~e \in E^{\text{mod}}, e.v_{\text{c}} = v\}$

\EndProcedure
\vspace{0.5em} 
\Procedure{MinWeightMatching}{$(V_{\text{c}}, V_{\text{np}}, E)$}
    \State \textbf{Input:} $(V_{\text{c}}, V_{\text{np}}, E)$: Vertex subsets and edges of a bipartite graph
    \State \textbf{Output:} $(V_{\text{c}}^{\text{min}}, V_{\text{np}}^{\text{min}}, E^{\text{min}})$: Vertex subsets and edges for the min-weight matching in the input bipartite graph
    \LineComment{Can be found using the Hungarian method \cite{Kuhn1955}}
\EndProcedure

\end{algorithmic}
\end{multicols}
\caption{Modified BD}\label{alg:maxmin1}\end{algorithm*}

\section{Results and Discussion}
\label{sec:results_and_discussion}


\subsection{Metrics and Methods}
\label{subsec:Metrics_Methods}

For each of the network topologies listed in Section \ref{subsec:net_topologies}, we analyze the minimum and median received EPR-pair rates, and the fairness of their allocations. The minimum EPR-pair rate reflects the guaranteed rate to each node pair in the network. The median EPR-pair rate reflects the rate to at least half of the node pairs.  We choose median over a mean to reduce impact from the exponential relationship between path length and transmission success probability.  We also analyze the normalized minimum EPR-pair rates, which enables a comparison of allocation strategies across network configurations.
Our normalization is relative to a baseline, which is the minimum EPR-pair rate assigned by Round Robin. It is used since it is a well-known and intuitive strategy. 

The fairness of EPR rate allocation is quantified by the Jain index \cite{Jain}. The Jain index for values $x_i, i=1,\ldots,r$, is: 
\begin{align} 
J(x_1, x_2, \ldots, x_r) &= \frac{(\sum_{i=1}^{r} x_i)^2}{r\sum_{i=1}^{r}x_{i}^2}.
\end{align} 
The maximum Jain index is unity when all $x_i$'s are equal. Hence, this indicates complete fairness. The minimum Jain index $\frac{1}{r}$ indicates a completely unfair resource allocation. In our case, for a network with $n$ nodes, $r$ is the number of node pairs $r = \kappa = n(n-1)/2$. The dependence of the Jain index on $n$ limits its use when comparing fairness between networks with different numbers of nodes. 
We also use the Jain index to analyze the importance of source node location by calculating it for the minimum EPR-pair rates assigned for each possible source node location. A smaller Jain index (lower-bounded by $1/n$) indicates increased source node location importance, since this is due to a larger variation in the minimum EPR-pair rate across the choices of source node location.

For simple and ILEC networks, we use the minimum and median EPR-pair rate, and the Jain index to analyze the performance of the different channel allocation strategies, for different choices of source node placement, and values of WSS loss. 
For the random Watts-Strogatz networks with a fixed WSS loss of 4dB, we analyze the impact of input parameters $n$, $k$, and $\beta$ on the minimum and median EPR-pair rates, and the Jain index using the modified LPT and BD approximation algorithms. We consider these  because, in the fixed networks, BD is the best strategy for suboptimal source locations, while LPT is the best strategy for the optimal source location. The minimum and median EPR-pair rates, and the Jain index, are calculated for the max-min-optimal source node for each of the two algorithms, and are averaged over 40 randomly generated Watt-Strogatz topologies for each value of input parameters $n$, $k$, and $\beta$ given in Section \ref{subsec:net_topologies}. 

\subsection{Spectrum Allocation in Simple and ILEC Networks}
\label{subsec:Fixed_Networks}

\begin{figure}[!th]
  \centering
  \subfloat[Minimum received EPR-pair rates]{\includegraphics[width=\columnwidth]{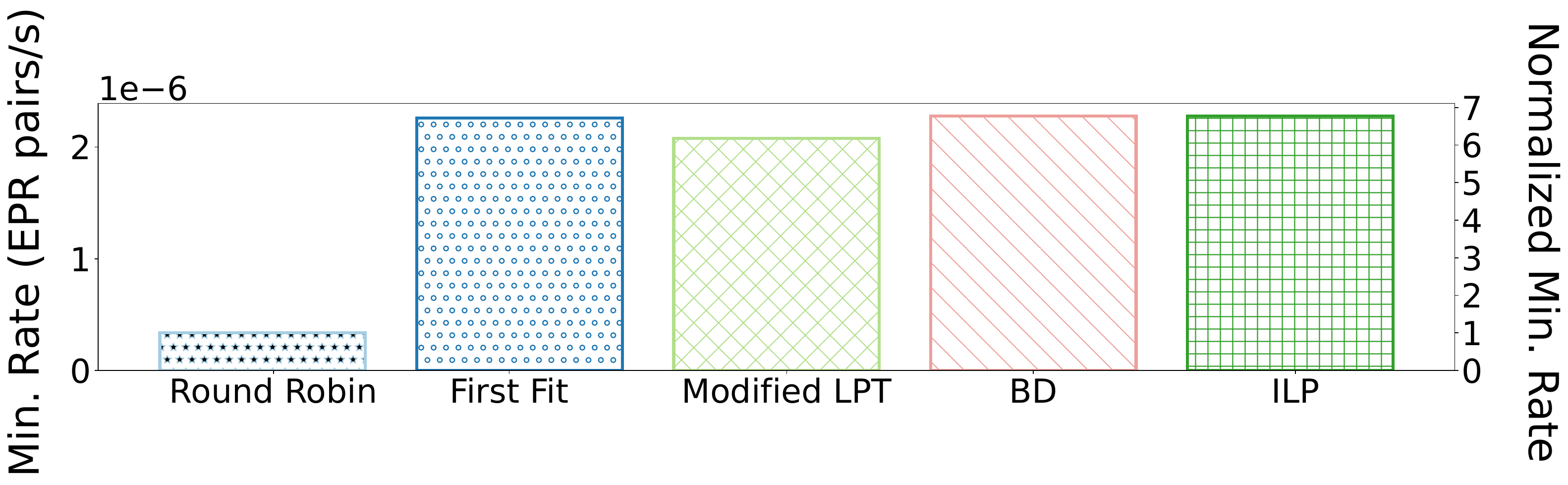} \label{fig:simple_results}}
  
  \subfloat[Median received EPR-pair rates]{\includegraphics[width=\columnwidth]{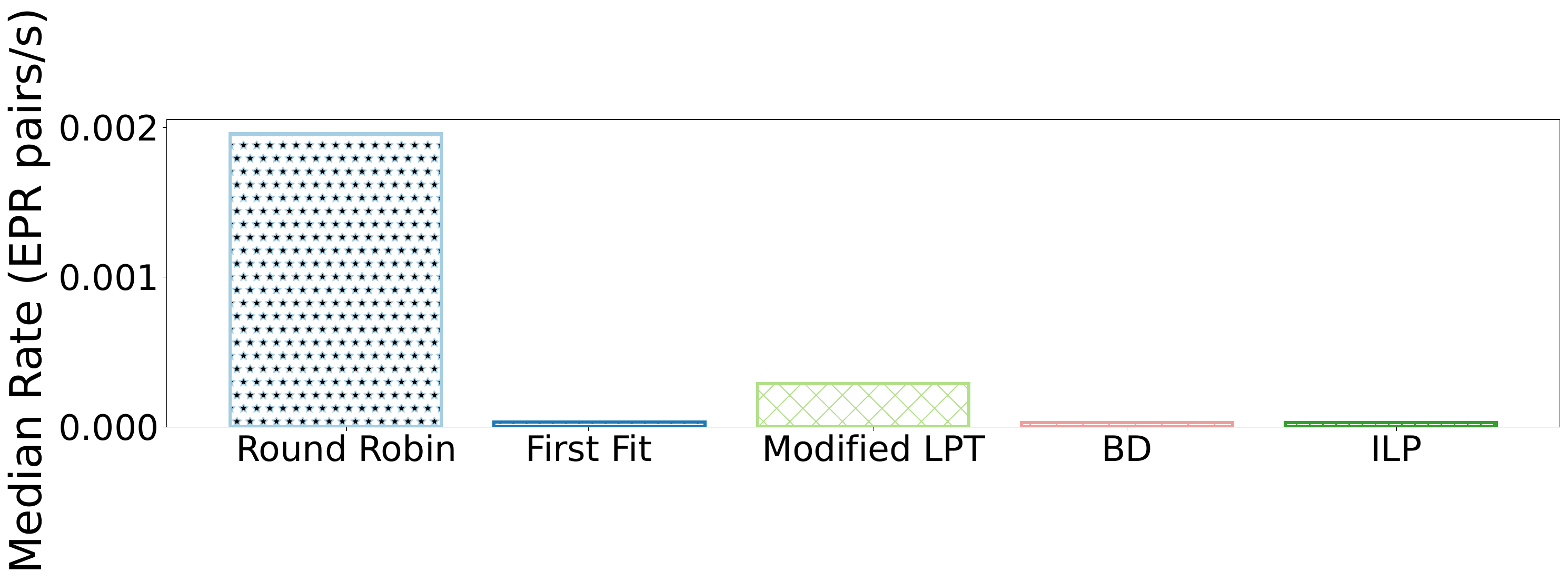} \label{fig:simple_results_median}}
  
  \subfloat[Jain index]{\includegraphics[width=\columnwidth]{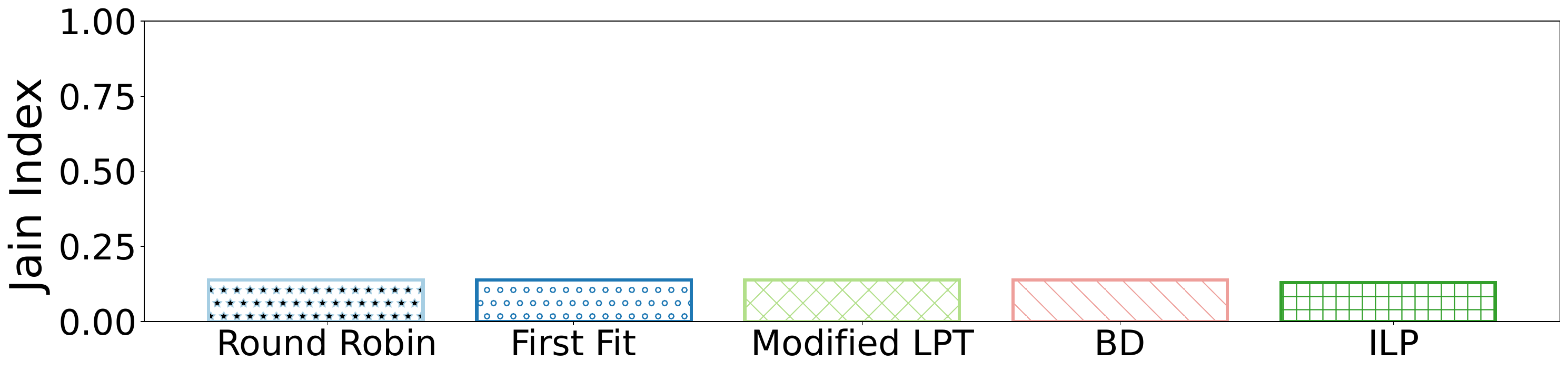} \label{simple_jain}}
  \caption{Comparison of performance using different allocation strategies on the simple network depicted in Fig.~\ref{fig:simple_layout} for 8 dB WSS loss and source node location A. In \protect\subref{fig:simple_results}, we report unnormalized (left-axis label) and normalized (right-axis label) minimum EPR-pair rates received by any node pair. \label{fig:simple}}
  \vspace{-0.2in}
\end{figure}

Fig.~\ref{fig:simple_results} depicts the minimum EPR-pair rates received by any node pair in the simple network topology depicted in Fig.~\ref{fig:simple_layout} when placing the source at node A and a WSS loss of 8 dB. We can calculate the optimal max-min solution using ILP for this configuration. We note that the BD and First Fit algorithms are close to optimal; the First Fit's performance is surprising given its relative simplicity. Modified LPT algorithm performs well, while Round Robin performs poorly. 

Fig.~\ref{fig:simple_results_median} depicts the median EPR-pair rates for the simple topology. ILP, which yields the maximal minimum EPR-pair rate, preforms poorly on this metric, as do the BD and First Fit algorithms. Round Robin, which produces the worst minimum EPR-pair rate, performs best here, followed by modified LPT.

Fig.~\ref{simple_jain} examines the Jain index for the simple topology. Although the ILP algorithm consistently produces the same minimum EPR-pair rate in each run, it may use distinct assignment configurations, resulting in varying Jain index. Thus, the results from the ILP algorithm are averaged over 1000 runs, with each run randomizing the order of processing the node pairs. The confidence intervals are negligibly small and are not reported. All strategies perform comparably. 

The performance variation of channel assignment algorithms across these metrics emphasizes their inherent limitations and underscores the importance of considering them collectively. While the minimum EPR-pair rate provides insights into the worst-case performance, it may not be representative of the majority of the rate allocations. Similarly, the median overlooks outliers in the allocation distribution. Additionally, the Jain index assesses the relative fairness among assignments but does not consider the magnitude of EPR-pair rates allocated.

\begin{figure*}[!th]
  \centering
   \subfloat{\raisebox{1\height}{} \includegraphics[width=\columnwidth]{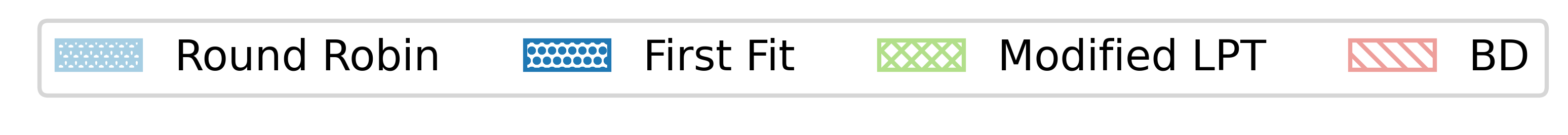}}
   
  \addtocounter{subfigure}{-1}
  \subfloat[Minimum received EPR-pair rates]{\includegraphics[width=1.99\columnwidth]{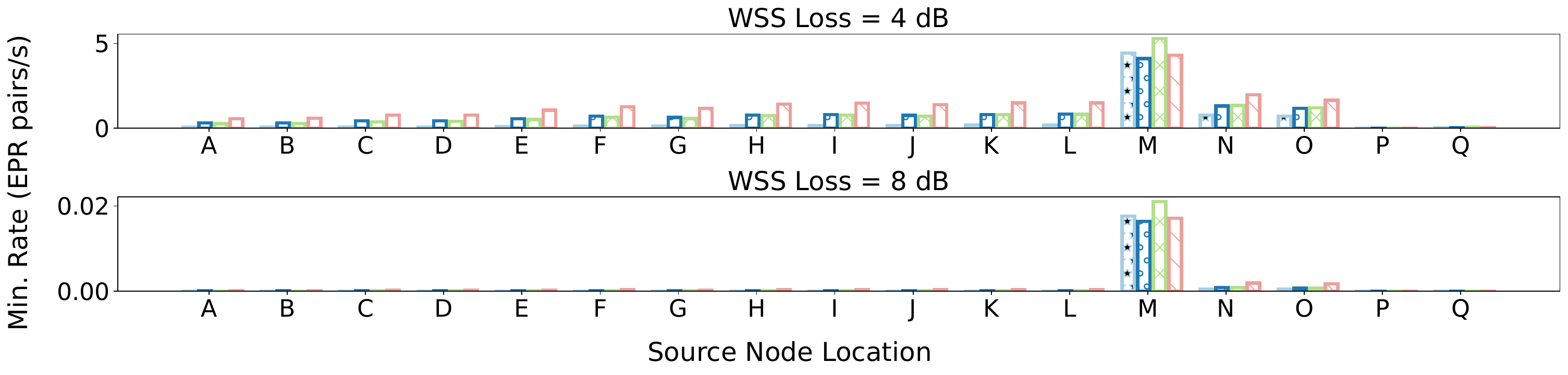}\label{fig:manhattan_results}}

    \subfloat[Normalized Minimum received EPR-pair rates]{\includegraphics[width=1.99\columnwidth]{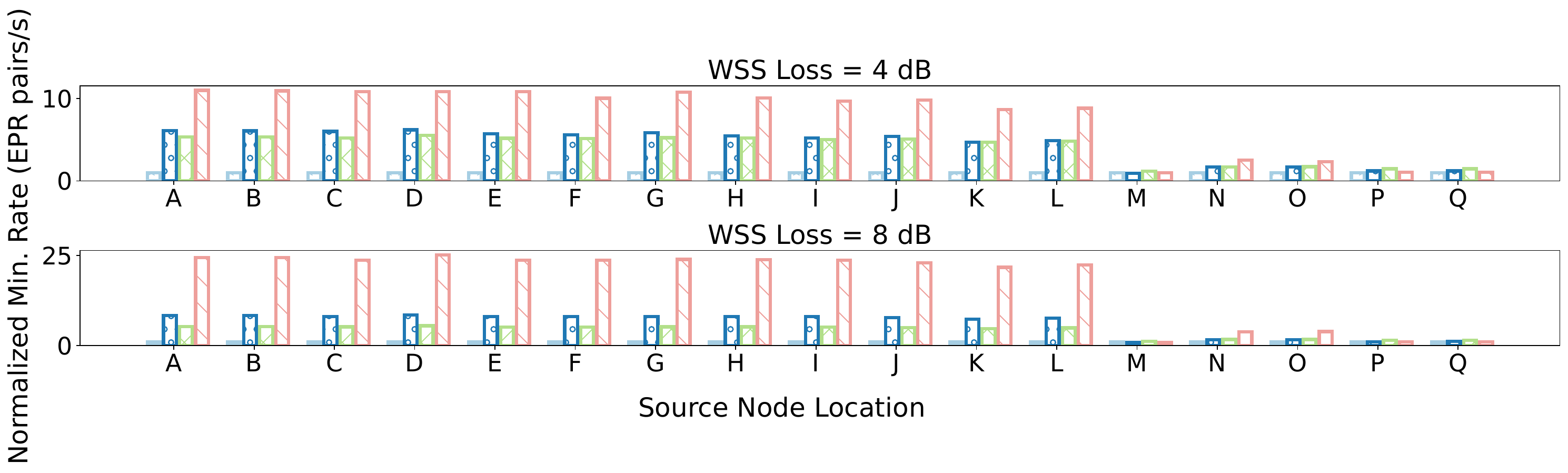}\label{fig:manhattan_results_norm}}

    \subfloat[Median received EPR-pair rates]{\includegraphics[width=1.99\columnwidth]{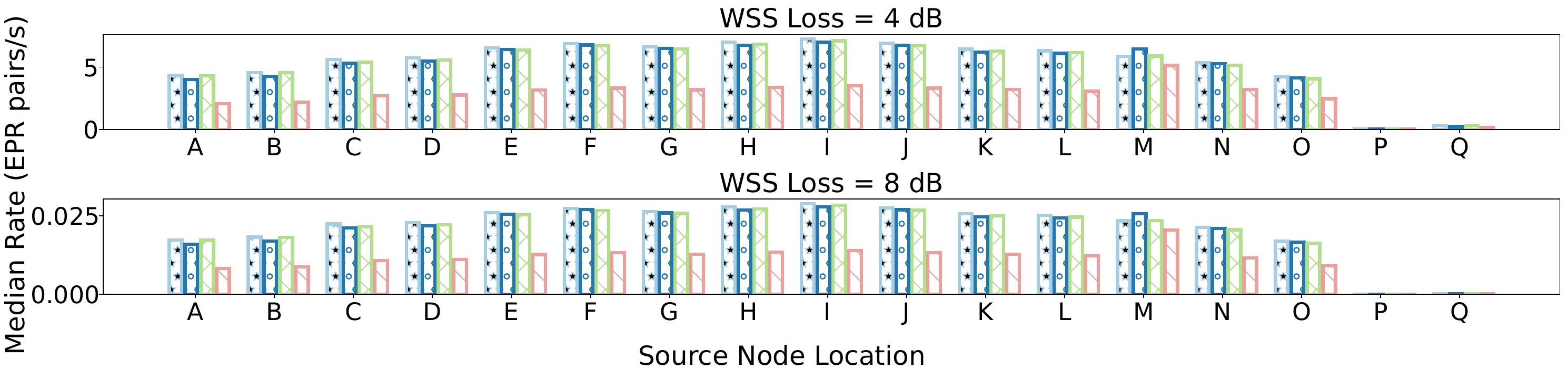}\label{fig:manhattan_results_median}}

  \subfloat[Jain index]{\includegraphics[width=1.99\columnwidth]{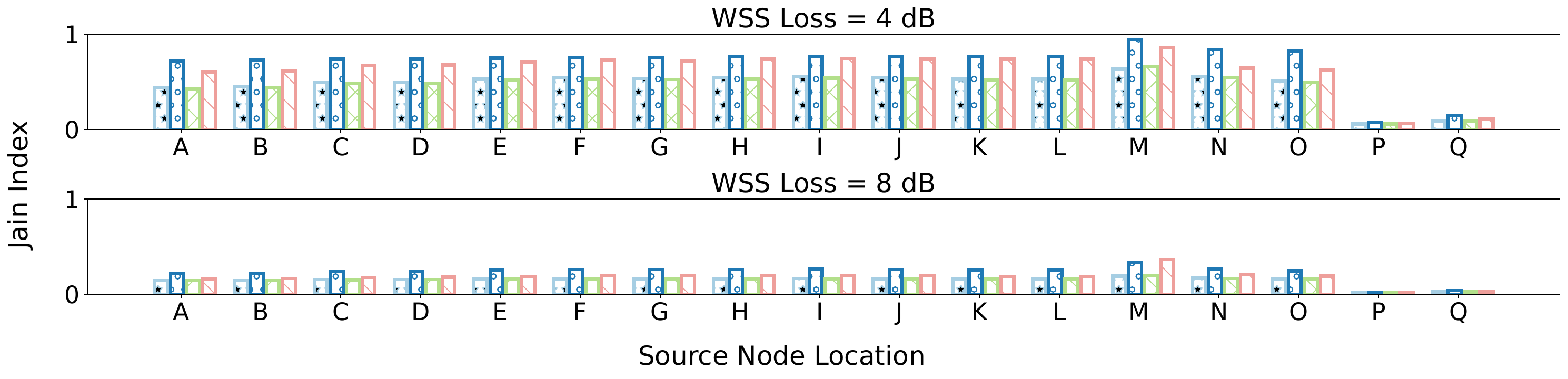} \label{fig:manhattan_jain}}
  \caption{Comparison of performance using different allocation strategies on the ILEC network depicted in Fig.~\ref{fig:manhattan_layout} for 4 dB and 8 dB WSS loss.\label{fig:manhattan}}
  \vspace{0.2in}
\end{figure*}

Figs.~\ref{fig:manhattan_results} and ~\ref{fig:manhattan_results_norm} show the normalized and unnormalized minimum received EPR-pair rates in the ILEC network depicted in Fig.~\ref{fig:manhattan_layout} for source at different network nodes. Due to the complexity of the ILP program for this topology, we cannot calculate the optimal solution.
The number of intermediate nodes traversed by a path in the ILEC network varies significantly based on the source node location. A linear increase in the number of intermediate nodes traversed leads to an exponential decrease of the path transmittance. 
Hence, we observe that the optimal placement of the source is at node M, which is the only fully-connected node in the ILEC network. Placing the source at nodes N and O, which connect to all but one node carries a substantial penalty in max-min rate, due the need for a two-hop connection to node P. Other placements fare significantly worse due to the required two-hop connections to more than one node.
Also due to this exponential relationship between path loss in dB and transmittance, the difference in the minimum EPR-pair rates across source node locations is accentuated when the WSS loss is set to 8dB versus 4 dB. 

The normalized values for the minimum EPR-pair rate indicate the relative performance improvement of a strategy compared to a trivial (Round Robin) spectrum allocation strategy. All four polynomial-time algorithms described in Section \ref{sec:allocation_strategies} perform well when the source is placed optimally, at node M. Surprisingly, modified LPT algorithm surpasses the others in this scenario (as well when source is at nodes P and Q). On the other hand, BD algorithm outperforms the others for other placements: slightly when the source is at nodes N and O, and significantly when it is at nodes A--L. 
This indicates that a more careful approach is needed when the received EPR-pair rates vary substantially from node-to-node.

Consistent with findings for the simple network in Fig.~\ref{fig:simple_layout}, First-Fit, modified LPT, and BD algorithms are effective in optimizing the minimum received EPR-pair rates by node pairs in the ILEC network in Fig.~\ref{fig:manhattan_layout}. Performance of all the strategies (including Round-Robin) on the fully-connected node M is comparable. BD algorithm outperforms others at all source-node locations except at nodes P and Q, where the modified LPT algorithm performs best.  



Fig.~\ref{fig:manhattan_results_median} depicts the median EPR-pair rate in the ILEC network for the source at different network nodes. The median EPR-pair rates resulting from the channel assignment algorithms are more similar than the minimum EPR-pair rates. The BD algorithm, which demonstrates superior performance for the minimum EPR rate, does not perform as well as Round Robin and LPT on the median EPR rate metric.

Finally, Fig.~\ref{fig:manhattan_jain} examines the Jain index for ILEC topology. 
The First Fit algorithm performs best on this metric.  BD is close second in every scenario except  when WSS loss is 8 dB and the source is at node M, where it outperforms First Fit.

\begin{figure}[!t]
\includegraphics[width=\columnwidth]{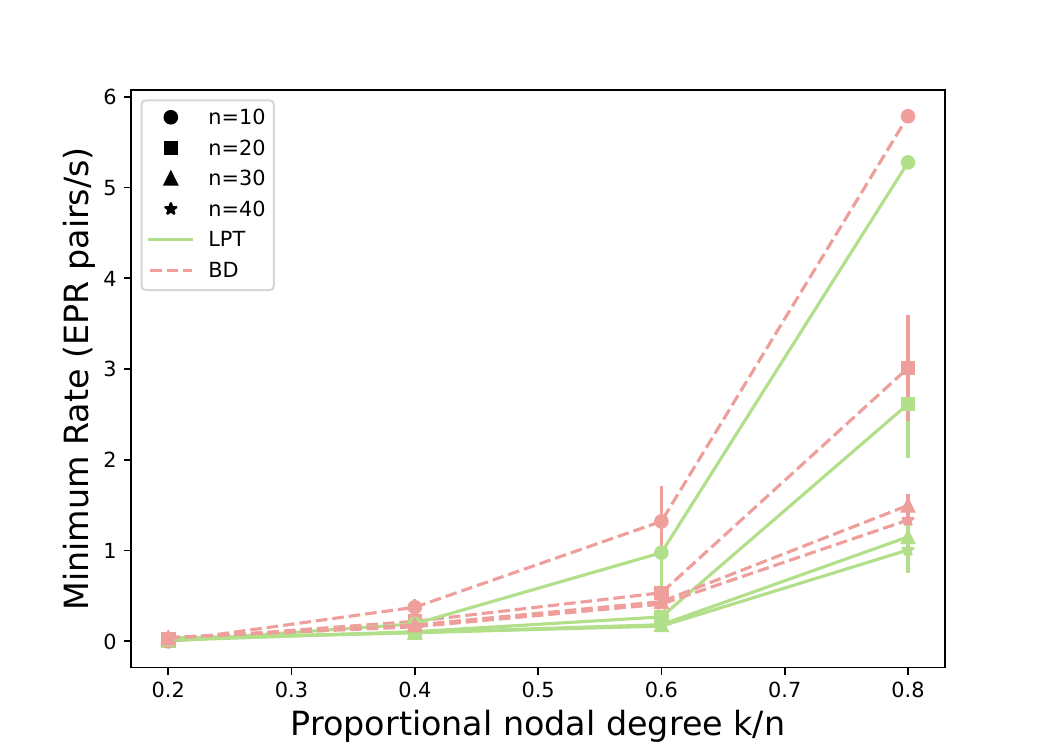}
\caption{Minimum EPR-pair rates at the optimal source location for Watts - Strogatz networks with varying number of nodes and proportional nodal degree, reported with 95\% confidence intervals. Rewiring probability $\beta=0.5$; results for other settings of $\beta$ are similar.}

\label{fig:ws_maxmin}
\end{figure}

\begin{figure*}[!t]
\centering
\subfloat[$n=10$, minimum Jain index = 0.1]{\includegraphics[width=\columnwidth]{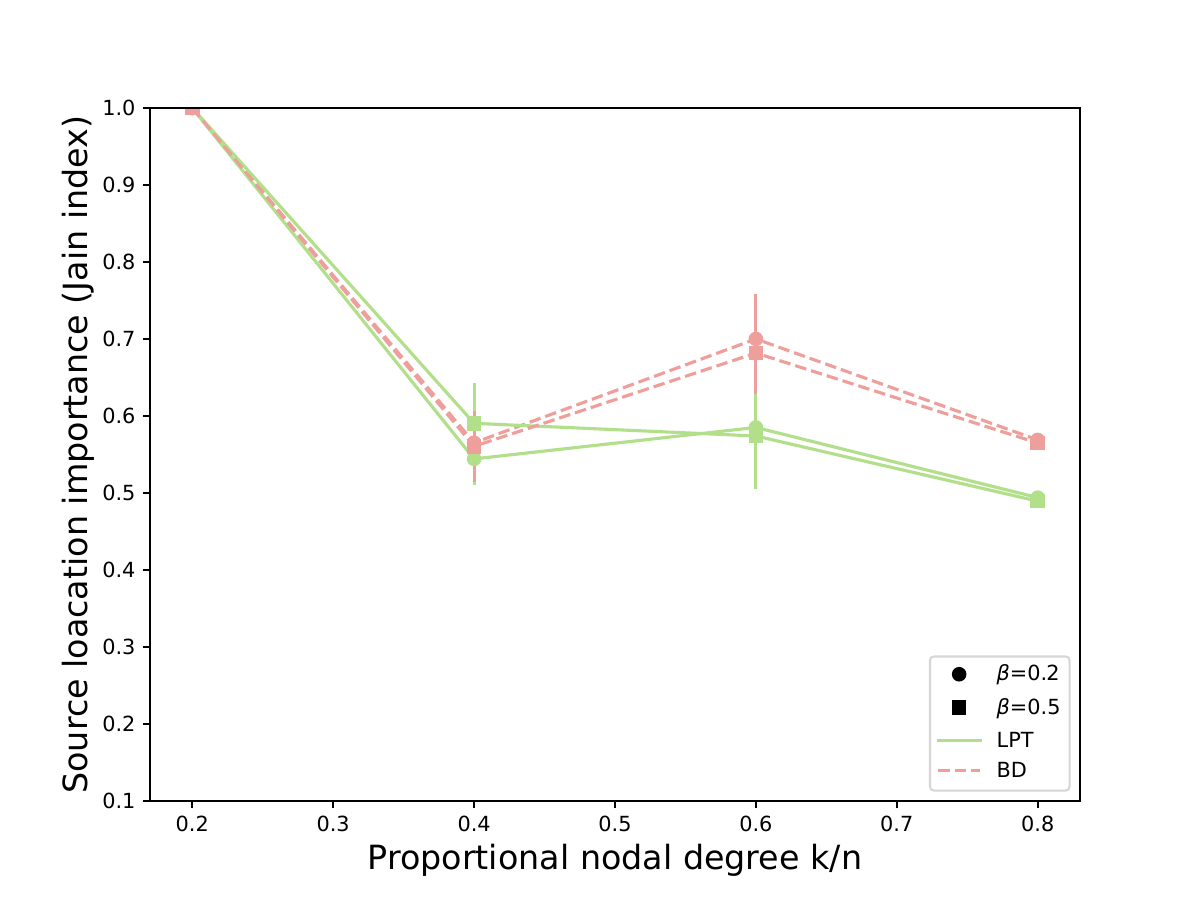} \label{fig:ws_jain_maxmin_10}}
\subfloat[$n=20$, minimum Jain index = 0.05]{\includegraphics[width=\columnwidth]{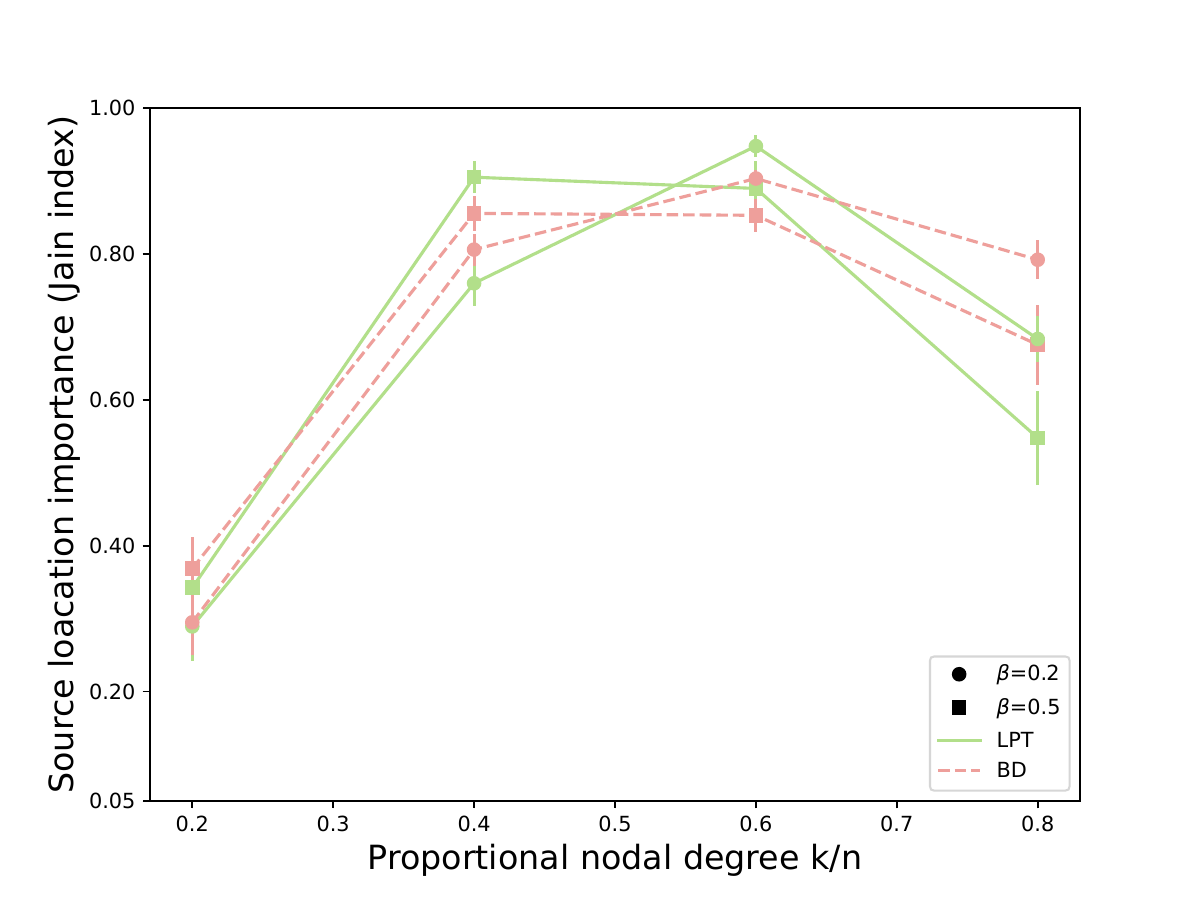} \label{fig:ws_jain_maxmin_20}}\\
\subfloat[$n=30$, minimum Jain index = 0.0333]{\includegraphics[width=\columnwidth]{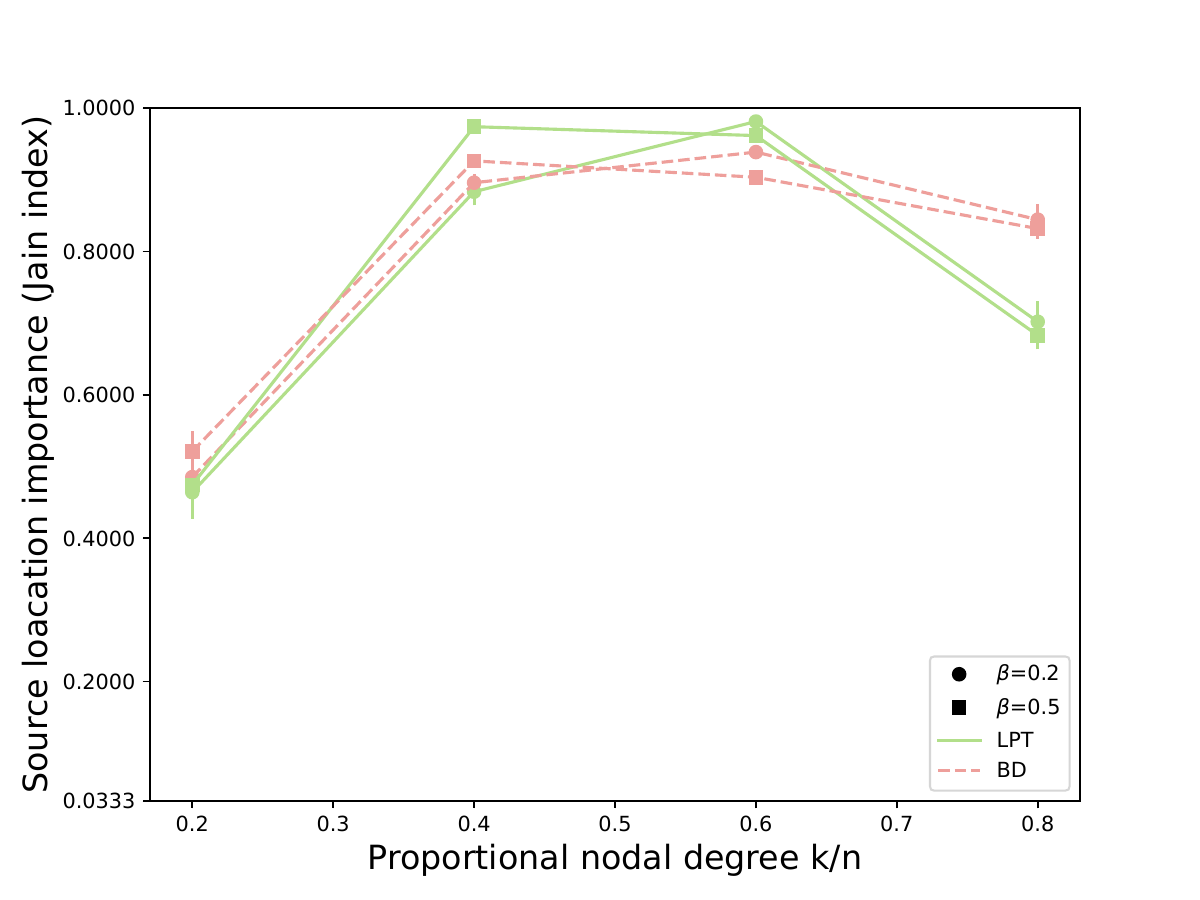} \label{fig:ws_jain_maxmin_30}}
\subfloat[$n=40$, minimum Jain index = 0.025]{\includegraphics[width=\columnwidth]{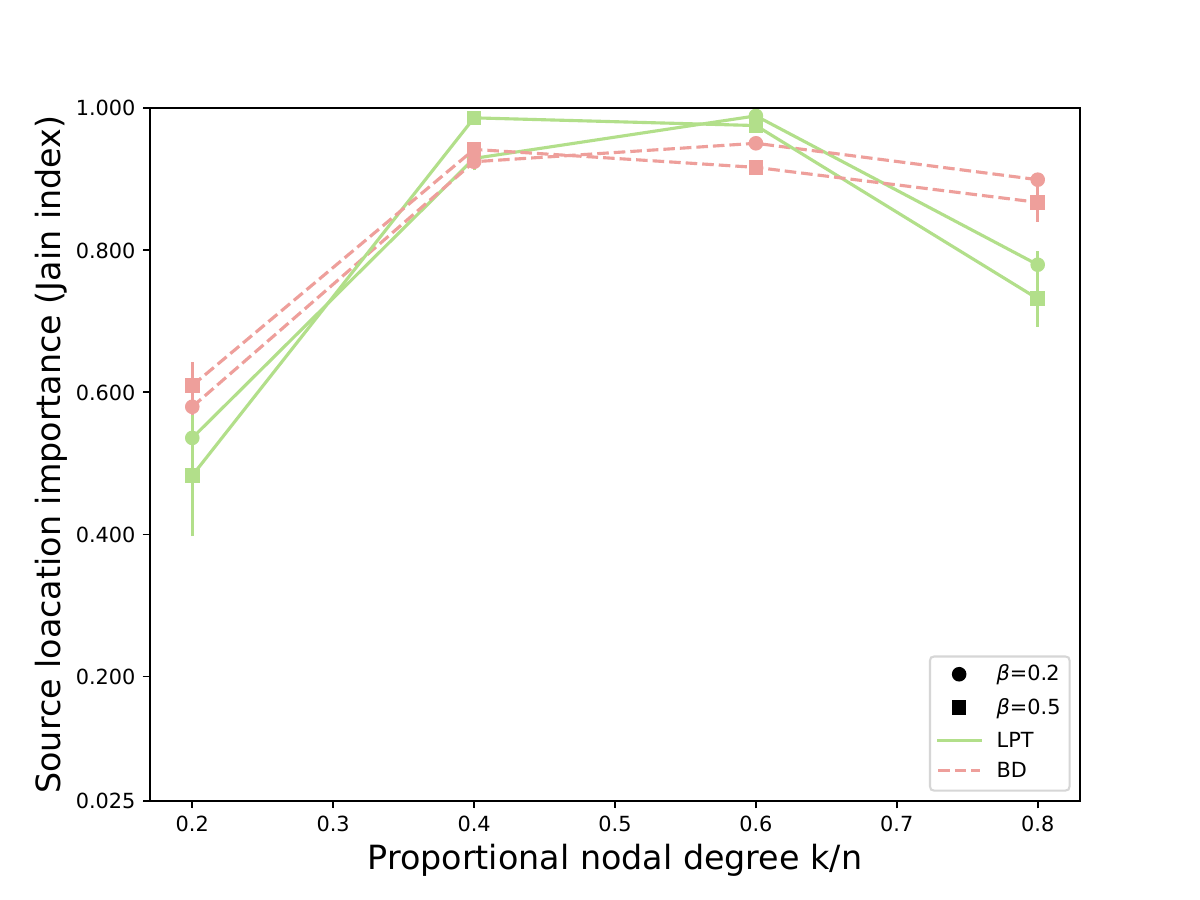} \label{fig:ws_jain_maxmin_40}}
\caption{Source node location importance measured on varying Watts-Strogatz networks using the Jain index for varying number of nodes, proportional nodal degree, and rewiring probability. Smaller Jain index indicates greater importance.}
\label{fig:ws_jain_maxmin}
\end{figure*}


\begin{figure*}[!t]
\centering
\subfloat[$n=10$, minimum Jain index = 0.0222]{\includegraphics[width=\columnwidth]{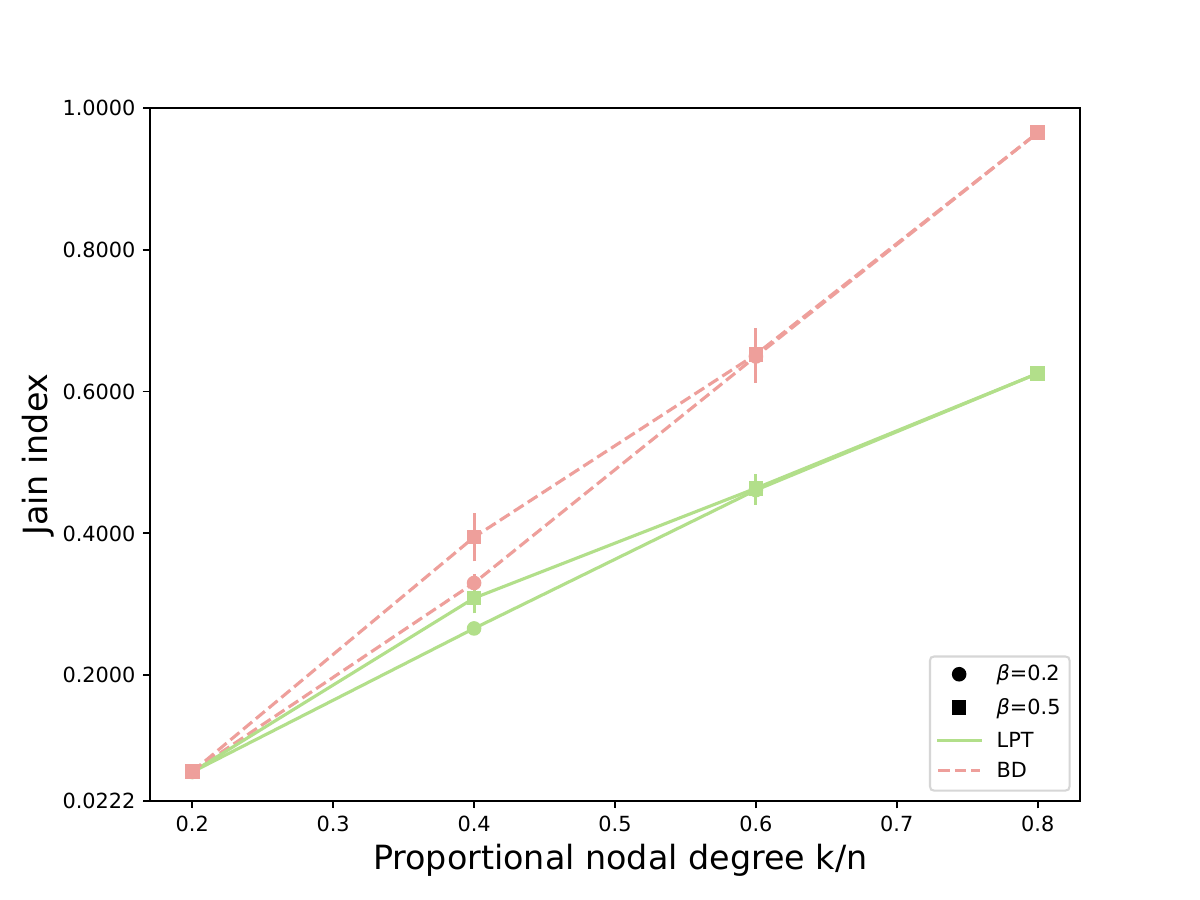} \label{fig:ws_jain_10}}
\subfloat[$n=20$, minimum Jain index = 0.0053]{\includegraphics[width=\columnwidth]{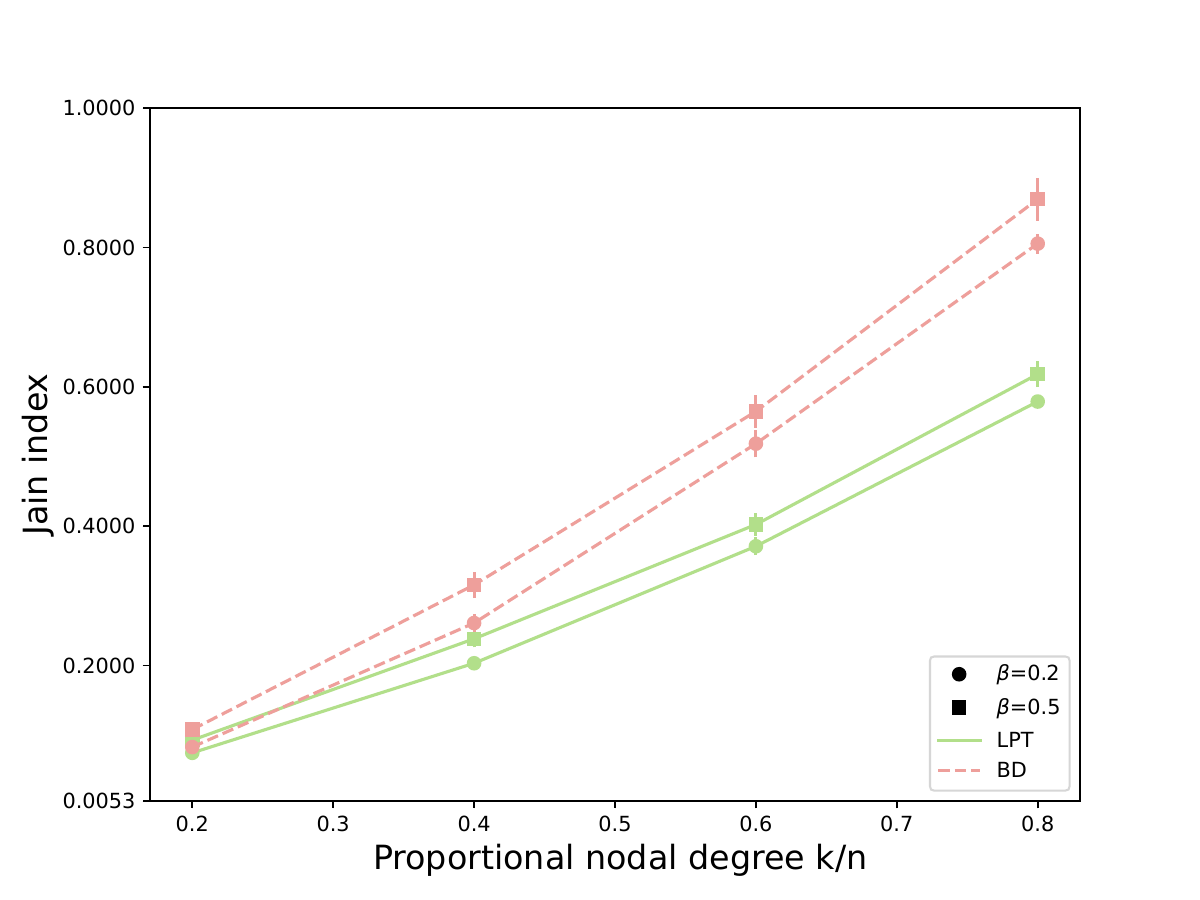} \label{fig:ws_jain_20}}\\
\subfloat[$n=30$, minimum Jain index = 0.0023]{\includegraphics[width=\columnwidth]{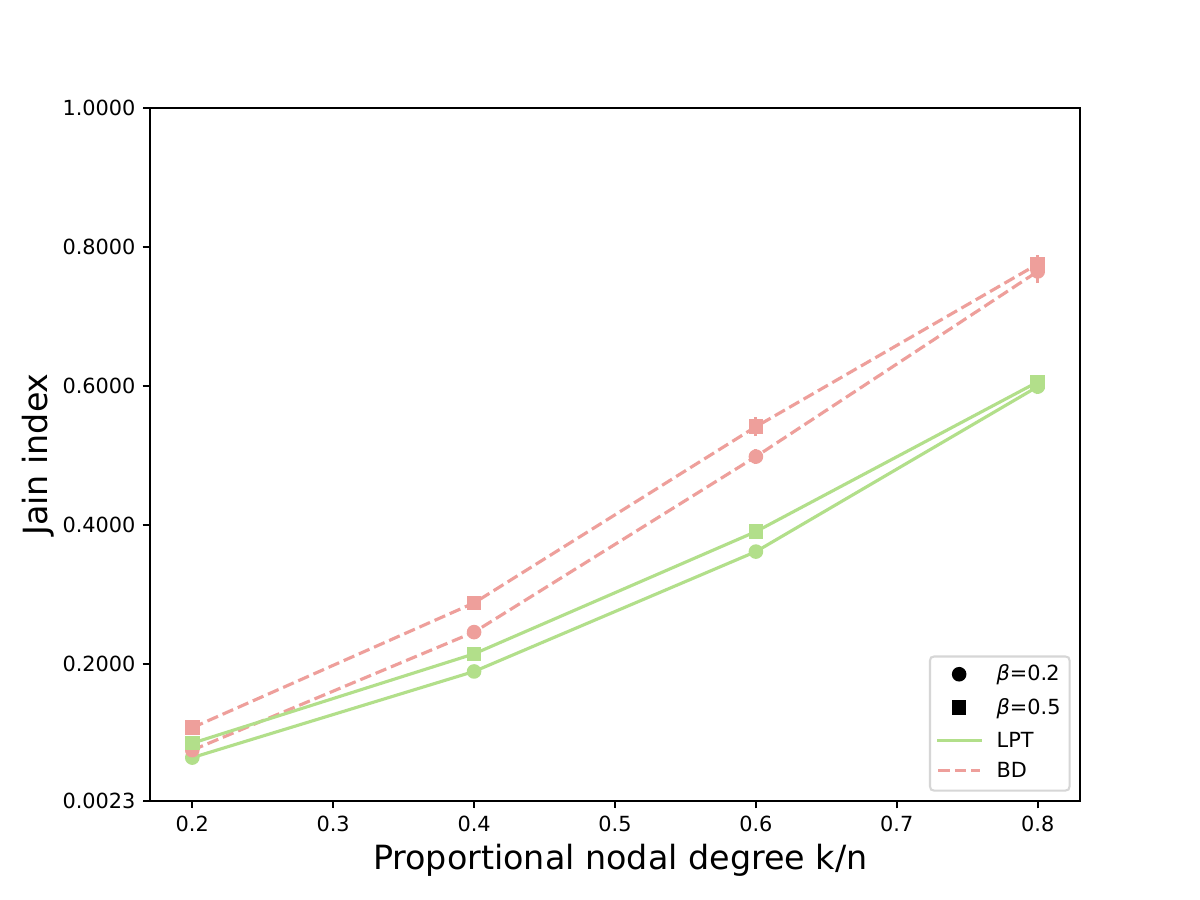} \label{fig:ws_jain_30}}
\subfloat[$n=40$, minimum Jain index = 0.0013]{\includegraphics[width=\columnwidth]{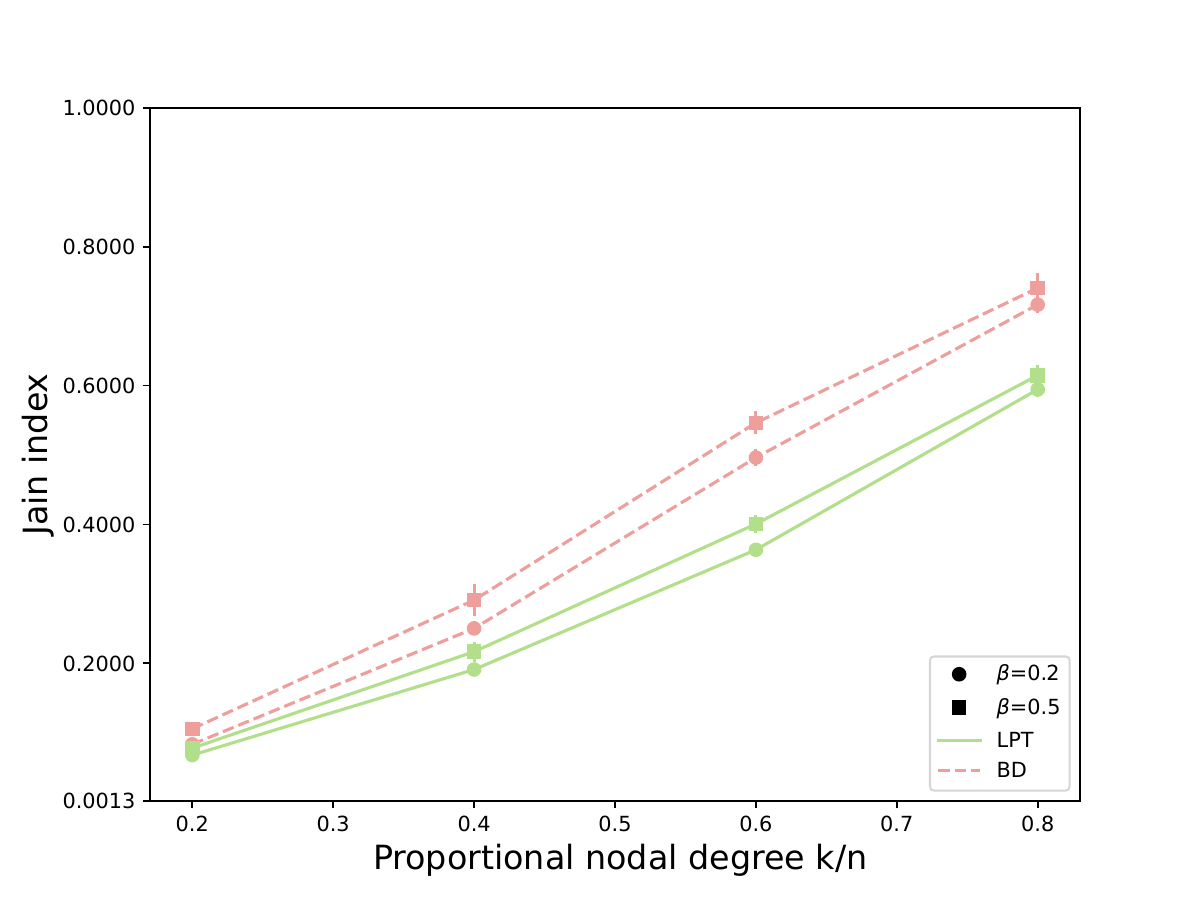} \label{fig:ws_jain_40}}
\caption{The fairness of the network and allocation strategy, measured on varying Watts-Strogatz networks using the Jain index at the node with the best max-min EPR-pair rate. Each subplot starts from its associated theoretical minimum value.}
\label{fig:ws_jain_value}
\end{figure*}

\begin{figure}[!t]
\centering
\includegraphics[width=\columnwidth]{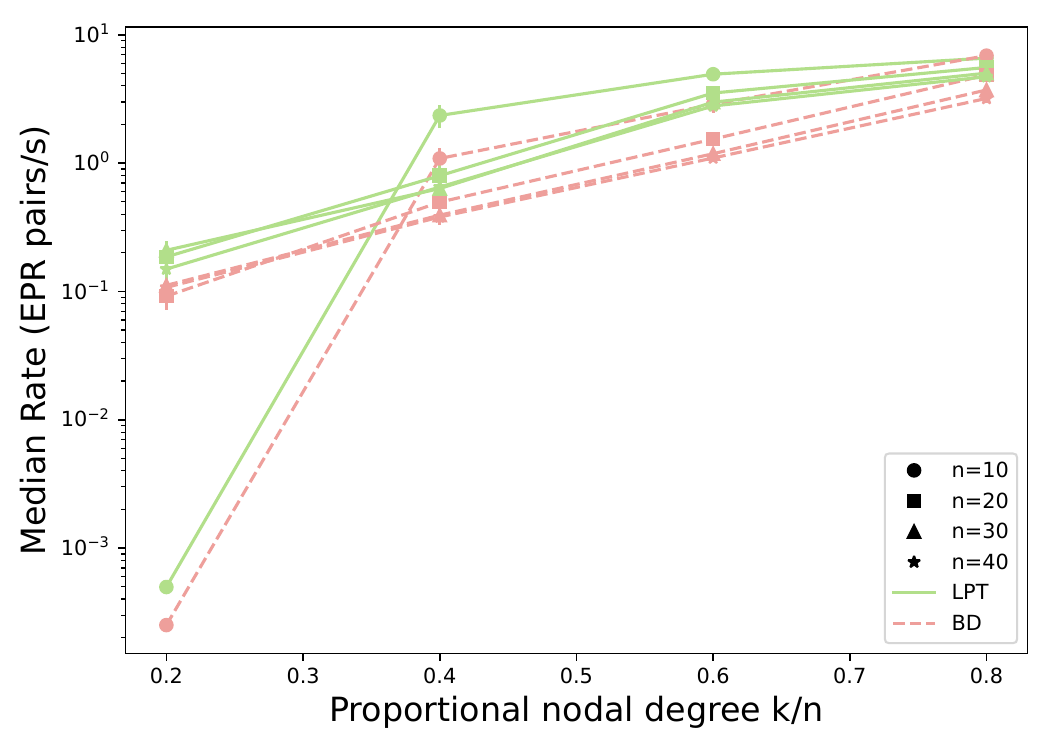}
\caption{Median EPR-pair rates at the max-min-optimal source location for Watts-Strogatz networks with varying number of nodes and proportional nodal degree. Rewiring probability $\beta=0.5$; results for other settings of $\beta$ are similar. Median EPR-pair rates on the ordinate are in log scale.}
\label{fig:ws_median_value}
\end{figure}


\subsection{Impact of Network Size and Connectivity}
\label{subsec:Topology_Dynamics}

 We analyze the impact of network size and connectivity using the Watts-Strogatz random networks described in Section \ref{subsec:net_topologies}. We focus on modified LPT and BD approximation algorithms as they perform well or better than others. Fig.~\ref{fig:ws_maxmin} shows that, as the number of nodes in a network increases, the minimum EPR-pair rate decreases. This is intuitive, as we expect that, with more nodes, the longer path lengths increase loss and decrease the likelihood that a photon successfully reaches its destination. Furthermore, as the proportional nodal degree $k/n$ increases, so does the minimum received EPR-pair rate. This is because increasing the number of connections in a network increases the possible available routing paths, thus, on average, decreasing the path lengths and loss in the best available paths. Finally, the rewiring probability $\beta$ does not seem to impact the max-min EPR-pair rate.  

Next, we study the importance of the source-node location on the max-min received EPR-pair rate using the Jain index. Larger Jain index implies that the source-node location is less important. Note that, for $n=10$ and $k=2$, varying the source-node location does not impact the max-min EPR-pair rate because the only topology generated using this $n$ and $k$ meeting the minimum cut requirement mentioned in \ref{subsec:network_arch} corresponds to a rotationally-symmetric ring network, where each node is only connected to its neighbor. Fig.~\ref{fig:ws_jain_maxmin} shows that, as the proportional nodal degree $k/n$ increases to approximately half connectivity, the importance of source location decreases. This is because of increasing uniformity of network path lengths. However, as $k/n$ climbs further, the source node placement again increases in importance. This is because now there are some nodes that are fully or almost-fully connected, making them better source choices relative to others. Since higher rewiring probability $\beta$ leads to a higher probability of having fully connected nodes, a higher source location importance for increased $\beta$ at high $k/n$ further confirms this. We also note that this corresponds to our results the ILEC network, which has a relatively high $k/n\approx 0.76$ and the Jain index $0.58$ for source node location (taken over the best performing algorithm at each), indicating its importance. This is surprising, given the dissimilar nature of these networks.

Fig.~\ref{fig:ws_jain_value} shows that, as proportional nodal degree $k/n$ increases, the spectrum allocation becomes more fair, indicated by an increase in the Jain index. This is due to the increasing uniformity in network path lengths driven by increasing $k/n$. The same also drives the improvement in spectrum allocation fairness with increasing rewiring probability.

Finally, we compare the median EPR-pair rates for the source node location optimizing the performance of BD and modified LPT algorithms in Fig.~\ref{fig:ws_median_value}. Although these algorithms approximate an optimal max-min EPR-pair rate solution, median measures the impact of the choice of algorithm beyond the minimum rate. Consistent with minimum EPR-pair rate result in Fig.~\ref{fig:ws_maxmin}, the median EPR-pair rate decreases with increasing number of nodes, and increases with increasing proportional nodal degree $k/n$. The modified LPT algorithm outperforms the BD algorithm in median EPR-pair rate, except in smaller networks at high values of $k/n$. 

\section{Conclusion}
\label{sec:conclusion}
In this study, we explore the optimization of EPR-pair distribution in repeaterless quantum networks that employ a source-in-the-middle time-frequency-heralded architecture. Specifically, we develop an optimal routing algorithm for such network. We also investigate various approaches for max-min fair spectrum allocation that approximate the optimal NP-hard solution, discovering that the BD approximation and modified LPT algorithms outperform others in EPR-pair rate while being comparable to others in another measure of fairness, the Jain index. However, the choice between modified LPT and BD approximation algorithms depends on the availability of computing resources (modified LPT needs significantly less) and desired performance metrics (BD approximation yields better minimum EPR-pair rate, while modified LPT outperforms in median EPR-pair rate and Jain index). Finally, we study the source node placement in the network, and find that the optimal location also strongly depends on the desired performance metrics. Although, our work already provides the routing and spectrum allocation approaches for the near-term repeaterless network deployments, our future work will focus on algorithm refinement as well as numerical and physical experiments. We will consider optimizing the ZALM source for network performance as well as employing alternative EPR-pair generation sources such as four-wave mixing processes. 
Furthermore, having multiple EPR-pair sources in a network substantially complicates the routing and spectrum allocation problem. We will explore various approaches to address it.

\section*{Acknowledgment}

We thank Vikash Kumar, Vivek Vasan, Dmitrii Briantcev, Prajit Dhara, Kevin C.~Chen, J.~Gabriel Richardson, Dennis McNulty, Michael G.~Raymer, Brian J.~Smith, Clark Embleton, and Robert N.~Norwood for useful discussions on this work.  We are especially grateful to Jeffrey H.~Shapiro for answering our questions about \cite{shapiro2024entanglementsourcequantummemory}.
This material is based upon High Performance Computing (HPC) resources supported by the University of Arizona TRIF, UITS, and Research, Innovation, and Impact (RII) and maintained by the UArizona Research Technologies department.

\bibliographystyle{IEEEtran}

\bibliography{references}

\begin{thebibliography}{10}
\providecommand{\url}[1]{#1}
\csname url@samestyle\endcsname
\providecommand{\newblock}{\relax}
\providecommand{\bibinfo}[2]{#2}
\providecommand{\BIBentrySTDinterwordspacing}{\spaceskip=0pt\relax}
\providecommand{\BIBentryALTinterwordstretchfactor}{4}
\providecommand{\BIBentryALTinterwordspacing}{\spaceskip=\fontdimen2\font plus
\BIBentryALTinterwordstretchfactor\fontdimen3\font minus \fontdimen4\font\relax}
\providecommand{\BIBforeignlanguage}[2]{{%
\expandafter\ifx\csname l@#1\endcsname\relax
\typeout{** WARNING: IEEEtran.bst: No hyphenation pattern has been}%
\typeout{** loaded for the language `#1'. Using the pattern for}%
\typeout{** the default language instead.}%
\else
\language=\csname l@#1\endcsname
\fi
#2}}
\providecommand{\BIBdecl}{\relax}
\BIBdecl

\bibitem{ICC_2024}
R.~Bali, A.~Tittelbaugh, S.~L. Jenkins, A.~Agrawal, J.~Horgan, M.~Ruffini, D.~Kilper, and B.~A. Bash, ``Routing and spectrum allocation in broadband degenerate {EPR}-pair distribution,'' in \emph{Proc. {IEEE} Int. Conf. Commun. (ICC)}, Denver, CO, USA, Jun. 2024, pp. 4954--4960.

\bibitem{wehner2018quantum}
S.~Wehner, D.~Elkouss, and R.~Hanson, ``Quantum internet: A vision for the road ahead,'' \emph{Science}, vol. 362, no. 6412, p. eaam9288, 2018.

\bibitem{Pirandola2018}
\BIBentryALTinterwordspacing
S.~Pirandola, B.~E.~A. Saleh, T.~Gehring, C.~Weedbrook, and S.~Lloyd, ``Advances in photonic quantum sensing,'' \emph{Nat. Photon.}, vol.~12, pp. 724--733, 2018. [Online]. Available: \url{https://doi.org/10.1038/s41566-018-0301-6}
\BIBentrySTDinterwordspacing

\bibitem{quantumSensing}
B.~Kantsepolsky, I.~Aviv, R.~Weitzfeld, and E.~Bordo, ``Exploring quantum sensing potential for systems applications,'' \emph{IEEE Access}, vol.~11, pp. 31\,569--31\,582, 2023.

\bibitem{distributedQuantumComputing}
P.~Singh, R.~Dasgupta, A.~Singh, H.~Pandey, V.~Hassija, V.~Chamola, and B.~Sikdar, ``A survey on available tools and technologies enabling quantum computing,'' \emph{IEEE Access}, vol.~12, pp. 57\,974--57\,991, 2024.

\bibitem{PhysRevLett.67.661}
\BIBentryALTinterwordspacing
A.~K. Ekert, ``Quantum cryptography based on {Bell's} theorem,'' \emph{Phys. Rev. Lett.}, vol.~67, pp. 661--663, Aug 1991. [Online]. Available: \url{https://link.aps.org/doi/10.1103/PhysRevLett.67.661}
\BIBentrySTDinterwordspacing

\bibitem{Wengerowsky}
\BIBentryALTinterwordspacing
S.~Wengerowsky, S.~K. Joshi, F.~Steinlechner, H.~H{\"u}bel, and R.~Ursin, ``An entanglement-based wavelength-multiplexed quantum communication network,'' \emph{Nature}, vol. 564, no. 7735, pp. 225--228, 2018. [Online]. Available: \url{https://doi.org/10.1038/s41586-018-0766-y}
\BIBentrySTDinterwordspacing

\bibitem{Zhu}
\BIBentryALTinterwordspacing
E.~Y. Zhu, C.~Corbari, A.~Gladyshev, P.~G. Kazansky, H.-K. Lo, and L.~Qian, ``Toward a reconfigurable quantum network enabled by a broadband entangled source,'' \emph{J. Opt. Soc. Am. B}, vol.~36, no.~3, pp. B1--B6, Mar. 2019. [Online]. Available: \url{https://opg.optica.org/josab/abstract.cfm?URI=josab-36-3-B1}
\BIBentrySTDinterwordspacing

\bibitem{Wang}
R.~Wang, O.~Alia, M.~J. Clark, S.~Bahrani, S.~K. Joshi, D.~Aktas, G.~T. Kanellos, M.~Peranić, M.~Lončarić, M.~Stipčević, J.~Rarity, R.~Nejabati, and D.~Simeonidou, ``A dynamic multi-protocol entanglement distribution quantum network,'' in \emph{2022 Opt. Fiber Commun. Conf. Exhib. (OFC)}, 2022, pp. 1--3.

\bibitem{alshowkan2024}
\BIBentryALTinterwordspacing
M.~Alshowkan, J.~M. Lukens, H.-H. Lu, and N.~A. Peters, ``Resilient entanglement distribution in a multihop quantum network,'' 2024. [Online]. Available: \url{https://arxiv.org/abs/2407.20443}
\BIBentrySTDinterwordspacing

\bibitem{pant2019routing}
M.~Pant, H.~Krovi, D.~Towsley, L.~Tassiulas, L.~Jiang, P.~Basu, D.~Englund, and S.~Guha, ``Routing entanglement in the quantum internet,'' \emph{NPJ Quantum Inf.}, vol.~5, no.~1, p.~25, 2019.

\bibitem{Li2024Survey}
Y.~Li, H.~Zhang, C.~Zhang, T.~Huang, and F.~R. Yu, ``A survey of quantum internet protocols from a layered perspective,'' \emph{{IEEE} Commun. Surv. Tutor.}, pp. 1--1, 2024.

\bibitem{Chen2024Fairness}
L.~Chen, K.~Xue, J.~Li, Z.~Li, R.~Li, N.~Yu, Q.~Sun, and J.~Lu, ``Redp: Reliable entanglement distribution protocol design for large-scale quantum networks,'' \emph{J. Sel. Areas Commun. ({JSAC})}, vol.~42, no.~7, pp. 1723--1737, 2024.

\bibitem{Yang2022Fairness}
L.~Yang, Y.~Zhao, H.~Xu, and C.~Qiao, ``Online entanglement routing in quantum networks,'' in \emph{Proc. IEEE/ACM Int. Symp. Qual. Service (IWQoS)}, 2022, pp. 1--10.

\bibitem{Zhao2023mixed}
G.~Zhao, J.~Wang, Y.~Zhao, H.~Xu, L.~Huang, and C.~Qiao, ``Segmented entanglement establishment with all-optical switching in quantum networks,'' \emph{IEEE/ACM Trans. Netw.}, vol.~32, no.~1, pp. 268--282, 2024.

\bibitem{muralidharan2016optimal}
S.~Muralidharan, L.~Li, J.~Kim, N.~L{\"u}tkenhaus, M.~D. Lukin, and L.~Jiang, ``Optimal architectures for long distance quantum communication,'' \emph{Sci. Rep.}, vol.~6, no.~1, p. 20463, 2016.

\bibitem{wang2022pre}
Y.~Wang, X.~Yu, Y.~Zhao, A.~Nag, and J.~Zhang, ``Pre-established entanglement distribution algorithm in quantum networks,'' \emph{{IEEE} J. Opt. Commun. Netw. (JOCN)}, vol.~14, no.~12, pp. 1020--1033, 2022.

\bibitem{patil2022entanglement}
A.~Patil, M.~Pant, D.~Englund, D.~Towsley, and S.~Guha, ``Entanglement generation in a quantum network at distance-independent rate,'' \emph{NPJ Quantum Inf.}, vol.~8, no.~1, p.~51, 2022.

\bibitem{panigrahy2023scalable}
N.~K. Panigrahy, M.~G. De~Andrade, S.~Pouryousef, D.~Towsley, and L.~Tassiulas, ``Scalable multipartite entanglement distribution in quantum networks,'' in \emph{Proc. IEEE Int. Conf. of Quantum Comput. Eng. (QCE)}, vol.~2, Sydney, Australia, 2023, pp. 391--392.

\bibitem{kaur2023distribution}
E.~Kaur and S.~Guha, ``Distribution of entanglement in two-dimensional square grid network,'' in \emph{Proc. IEEE Int. Conf. Quantum Comput. Eng. (QCE)}, Sydney, Australia, 2023, pp. 1154--1164.

\bibitem{sutcliffe2023multi}
E.~Sutcliffe and A.~Beghelli, ``Multi-user entanglement distribution in quantum networks using multipath routing,'' \emph{{IEEE} Trans. Quantum Eng. (TQE)}, vol.~4, pp. 1--15, 2023.

\bibitem{van2023entanglement}
E.~A. Van~Milligen, E.~Jacobson, A.~Patil, G.~Vardoyan, D.~Towsley, and S.~Guha, ``Entanglement routing over networks with time multiplexed repeaters,'' \emph{arXiv preprint arXiv:2308.15028}, 2023.

\bibitem{Chen}
\BIBentryALTinterwordspacing
K.~C. Chen, P.~Dhara, M.~Heuck, Y.~Lee, W.~Dai, S.~Guha, and D.~Englund, ``Zero-added-loss entangled-photon multiplexing for ground- and space-based quantum networks,'' \emph{Phys. Rev. Appl.}, vol.~19, p. 054029, May 2023. [Online]. Available: \url{https://link.aps.org/doi/10.1103/PhysRevApplied.19.054029}
\BIBentrySTDinterwordspacing

\bibitem{shapiro2024entanglementsourcequantummemory}
\BIBentryALTinterwordspacing
J.~H. Shapiro, M.~G. Raymer, C.~Embleton, F.~N. Wong, and B.~J. Smith, ``Entanglement source and quantum memory analysis for zero-added-loss multiplexing,'' \emph{Phys. Rev. Appl.}, vol.~22, p. 044014, Oct 2024. [Online]. Available: \url{https://link.aps.org/doi/10.1103/PhysRevApplied.22.044014}
\BIBentrySTDinterwordspacing

\bibitem{Suurballe}
J.~Suurballe and R.~Tarjan, ``\BIBforeignlanguage{English (US)}{A quick method for finding shortest pairs of disjoint paths},'' \emph{\BIBforeignlanguage{English (US)}{Networks}}, vol.~14, no.~2, pp. 325--336, 1984.

\bibitem{Banerjee}
\BIBentryALTinterwordspacing
S.~Banerjee, R.~Ghosh, and A.~Reddy, ``Parallel algorithm for shortest pairs of edge-disjoint paths,'' \emph{J. Parallel Distrib. Comput.}, vol.~33, no.~2, p. 165–171, Mar. 1996. [Online]. Available: \url{https://doi.org/10.1006/jpdc.1996.0035}
\BIBentrySTDinterwordspacing

\bibitem{Simmons}
J.~M. Simmons, \emph{Optical network design and planning}.\hskip 1em plus 0.5em minus 0.4em\relax Springer, 2014.

\bibitem{Chatterjee}
B.~C. Chatterjee, N.~Sarma, and E.~Oki, ``Routing and spectrum allocation in elastic optical networks: A tutorial,'' \emph{IEEE Commun. Surv. Tutor.}, vol.~17, no.~3, pp. 1776--1800, 2015.

\bibitem{Saberi}
\BIBentryALTinterwordspacing
R.~J. Lipton, E.~Markakis, E.~Mossel, and A.~Saberi, ``On approximately fair allocations of indivisible goods,'' in \emph{Proc. ACM Conf. Electron. Commerce}, New York, NY, USA, 2004, p. 125–131. [Online]. Available: \url{https://doi.org/10.1145/988772.988792}
\BIBentrySTDinterwordspacing

\bibitem{Jain}
R.~K. Jain, D.-M.~W. Chiu, and W.~R. Hawe, ``A quantitative measurement of fairness and discrimination for resource allocation in shared computer system,'' \emph{Eastern Research Laboratory, Digital Equipment Corporation: Hudson, MA, USA}, vol.~2, 1984.

\bibitem{Yu}
J.~Yu, Y.~Li, M.~Bhopalwala, S.~Das, M.~Ruffini, and D.~C. Kilper, ``Midhaul transmission using edge data centers with split phy processing and wavelength reassignment for 5g wireless networks,'' in \emph{Proc. Int. Conf. Opt. Netw. Des. Model. (ONDM)}, Dublin, Ireland, 2018, pp. 178--183.

\bibitem{Li}
Y.~Li, M.~Bhopalwala, S.~Das, J.~Yu, W.~Mo, M.~Ruffini, and D.~C. Kilper, ``Joint optimization of bbu pool allocation and selection for c-ran networks,'' in \emph{Proc. Opt. Fiber Commun. Conf. Exhib. (OFC)}, San Diego, CA, USA, 2018, pp. 1--3.

\bibitem{WattsStrogatz}
\BIBentryALTinterwordspacing
D.~J. Watts and S.~H. Strogatz, ``Collective dynamics of `small-world'networks,'' \emph{Nature}, vol. 393, no. 6684, pp. 440--442, 1998. [Online]. Available: \url{https://doi.org/10.1038/30918}
\BIBentrySTDinterwordspacing

\bibitem{takeoka2014}
\BIBentryALTinterwordspacing
M.~Takeoka, S.~Guha, and M.~M. Wilde, ``Fundamental rate-loss tradeoff for optical quantum key distribution,'' \emph{Nat. Commun.}, vol.~5, no.~1, p. 5235, 2014. [Online]. Available: \url{https://doi.org/10.1038/ncomms6235}
\BIBentrySTDinterwordspacing

\bibitem{Pirandola2017}
\BIBentryALTinterwordspacing
S.~Pirandola, R.~Laurenza, C.~Ottaviani, and L.~Banchi, ``Fundamental limits of repeaterless quantum communications,'' \emph{Nat. Commun.}, vol.~8, no.~1, p. 15043, 2017. [Online]. Available: \url{https://doi.org/10.1038/ncomms15043}
\BIBentrySTDinterwordspacing

\bibitem{PhysRevLett.qunatumRepeater}
\BIBentryALTinterwordspacing
V.~Krutyanskiy, M.~Canteri, M.~Meraner, J.~Bate, V.~Krcmarsky, J.~Schupp, N.~Sangouard, and B.~P. Lanyon, ``Telecom-wavelength quantum repeater node based on a trapped-ion processor,'' \emph{Phys. Rev. Lett.}, vol. 130, p. 213601, May 2023. [Online]. Available: \url{https://link.aps.org/doi/10.1103/PhysRevLett.130.213601}
\BIBentrySTDinterwordspacing

\bibitem{LangenfeldQuantumRepeater}
\BIBentryALTinterwordspacing
S.~Langenfeld, P.~Thomas, O.~Morin, and G.~Rempe, ``Quantum repeater node demonstrating unconditionally secure key distribution,'' \emph{Phys. Rev. Lett.}, vol. 126, p. 230506, Jun 2021. [Online]. Available: \url{https://link.aps.org/doi/10.1103/PhysRevLett.126.230506}
\BIBentrySTDinterwordspacing

\bibitem{Lumentum}
\emph{TrueFlex® Twin High Port Count Wavelength Selective Switch (Twin WSS)}, 11 1997, revised Sept. 2002.

\bibitem{Lumentum_lowloss}
P.~D. Colbourne, S.~McLaughlin, C.~Murley, S.~Gaudet, and D.~Burke, ``Contentionless twin 8×24 wss with low insertion loss,'' in \emph{Proc. Opt. Fiber Commun. Conf. Exhib. (OFC)}, San Diego, CA, USA, 2018, pp. 1--3.

\bibitem{Kingfisher}
``Optical loss \& testing overview | {Kingfisher International.}'' \url{https://kingfisherfiber.com/application-notes/optical-loss-testing-overview}, accessed: Oct. 11, 2023.

\bibitem{Aziz}
\BIBentryALTinterwordspacing
H.~Aziz, I.~Caragiannis, A.~Igarashi, and T.~Walsh, ``Fair allocation of indivisible goods and chores,'' \emph{Auton. Agents and Multi-Agent Syst.}, vol.~36, no.~1, p.~3, 2021. [Online]. Available: \url{https://doi.org/10.1007/s10458-021-09532-8}
\BIBentrySTDinterwordspacing

\bibitem{Graham}
\BIBentryALTinterwordspacing
R.~L. Graham, ``Bounds on multiprocessing timing anomalies,'' \emph{SIAM J. Appl. Math.}, vol.~17, no.~2, pp. 416--429, 1969. [Online]. Available: \url{http://www.jstor.org/stable/2099572}
\BIBentrySTDinterwordspacing

\bibitem{Deuermeyer}
\BIBentryALTinterwordspacing
B.~L. Deuermeyer, D.~K. Friesen, and M.~A. Langston, ``Scheduling to maximize the minimum processor finish time in a multiprocessor system,'' \emph{SIAM J. Algebr. Discrete Methods}, vol.~3, no.~2, pp. 190--196, 1982. [Online]. Available: \url{https://doi.org/10.1137/0603019}
\BIBentrySTDinterwordspacing

\bibitem{Wu}
\BIBentryALTinterwordspacing
B.~Y. Wu, ``An analysis of the lpt algorithm for the max--min and the min--ratio partition problems,'' \emph{Theor. Comput. Sci.}, vol. 349, no.~3, pp. 407--419, 2005. [Online]. Available: \url{https://www.sciencedirect.com/science/article/pii/S0304397505005815}
\BIBentrySTDinterwordspacing

\bibitem{Bezakova}
\BIBentryALTinterwordspacing
I.~Bez\'{a}kov\'{a} and V.~Dani, ``Allocating indivisible goods,'' \emph{Proc. {ACM} {SIG}ecom}, vol.~5, no.~3, p. 11–18, apr 2005. [Online]. Available: \url{https://doi.org/10.1145/1120680.1120683}
\BIBentrySTDinterwordspacing

\bibitem{Tardos}
J.~Kleinberg and E.~Tardos, ``A first application: The bipartite matching problem,'' in \emph{Algorithm Design}.\hskip 1em plus 0.5em minus 0.4em\relax USA: Addison-Wesley Longman Publishing Co., Inc., 2005, pp. 367--373.

\bibitem{Kuhn1955}
H.~W. Kuhn, ``The hungarian method for the assignment problem,'' \emph{Nav. Res. Logistics Quart.}, vol.~2, no. 1-2, pp. 83--97, 1955.

\end{thebibliography}

\appendix
\label{Appendix}
Here we provide the details for the calculation of EPR-pair rates in Fig.~\ref{fig:dist_185}.
It is based on \cite{shapiro2024entanglementsourcequantummemory}, which derives \eqref{eq:wavefunction}--\eqref{eq:probability} and evaluates the heralding probability, heralding efficiency, and purity of the polarization-entangled photon pairs emitted from the ZALM EPR-pair source proposed in \cite{Chen}. We employ the SPDC parameters values in Table \ref{tab:param} in our calculations.

\begin{table}[h!]
\renewcommand{\arraystretch}{1.3}
\centering
\begin{threeparttable}
\caption{Table of parameters}
\label{tab:param}
\begin{tabular}{lll}
\hline
\textbf{Parameter} & \textbf{Symbol} & \textbf{Value} \\ 
\hline
Pump-pulse duration & $\sigma_{\text{P}}$ & 36 ps \\ 
Pump-pulse repetition rate & $r_{\text{P}}$ & $2.77\times10^{12}$  pulses/sec  \\ 
Phase-matching bandwidth\tnote{a} & $\Omega_{\text{PM}}/2\pi$ & 6.37 THz \\ 
Total bandwidth used & $B_{\text{s}}$ & 2.430 THz \\ 
Channel bandwidth & $B_{\text{c}}$ & 11 GHz \\ 
Channel spacing & $B_\Delta$ & 13.135 GHz \\ 
Number of channels & $m$ & 185 \\ 
\hline
\end{tabular}
\begin{tablenotes}
\item [a] We use the value for $\Omega_{\text{PM}}/2\pi$ from \cite[Table I]{shapiro2024entanglementsourcequantummemory}.
\end{tablenotes}
\end{threeparttable}
\end{table}

First, consider the all-Gaussian biphoton wave function \cite[Eq.~(9)]{shapiro2024entanglementsourcequantummemory}, 
\begin{equation}
    \begin{aligned}
        \Psi_{SI}(\omega_S,\omega_I) &= \sqrt{8\pi\sigma_{\text{P}}/\Omega_{\text{PM}}}e^{-(\Delta\omega_S+\Delta\omega_I)^2\sigma_{\text{P}}^2/16}\\
        &\phantom{=}\times \,e^{-4(\Delta\omega_S-\Delta\omega_I)^2/\Omega_{\text{PM}}^2},
    \end{aligned}
    \label{eq:wavefunction}
\end{equation}
where $\sigma_{\text{P}}$ is the pump-pulse duration of the SPDC pump signal, $\Omega_{\text{PM}}$ is the phase-matching bandwidth of the SPDC crystal, and $\Delta\omega_{c} \equiv \omega_{c} - \omega_{c_0} $, for $c = S,I$, are the signal and idler detunings from their respective center frequencies.
We assume $m$-channel ideal WDM filters for signal and idler, with odd $m$, and their frequency responses for channel $x$ being:
\begin{align}
    H_{S_x}(\omega_S)&=\left\{\begin{array}{ll}1,\left|\omega_S-\omega_{S_0}+2\pi\left(x-\frac{m+1}{2}\right) B_{\Delta}\right|\leq \pi B_{\text{c}}\\0,\text{otherwise,}\end{array}\right.\label{eq:signalfilter}\\
    H_{I_x}(\omega_I)&=\left\{\begin{array}{ll}1,\left|\omega_I-\omega_{I_0}-2\pi \left(x-\frac{m+1}{2}\right)B_{\Delta}\right|\leq \pi B_{\text{c}}\\0,\text{otherwise,}\end{array}\label{eq:idlerfilter}\right.
\end{align}
where $\omega_{S_0}$ and $\omega_{I_0}$ are signal and idler center frequencies that satisfy $\left|\omega_{S_0}-\omega_{I_0}\right|\gg\Omega_{\text{PM}}$ and $\omega_{S_0}+\omega_{I_0}=\omega_{\text{P}}$ for pump center frequency $\omega_{\text{P}}$. 
In this paper, $\omega_{S_0}=1.216\times10^{12}$ rad, corresponding to 1550 nm center wavelength of the C-band.
Assuming ideal photo-detection and using the wave function in \eqref{eq:wavefunction} and the ideal filter in \eqref{eq:idlerfilter}, we obtain the probability of generating a heralded EPR pair in channel $x$ \cite[Eq. (40)]{shapiro2024entanglementsourcequantummemory}:
\begin{align} \label{eq:probability}
    P_x = \frac{1}{4}\left(\int \frac{\dif\omega_S}{2\pi}\int \frac{\dif\omega_I}{2\pi} |\Psi_{S_xI_x}(\omega_S,\omega_I)|^2\right)^2,
\end{align}
where $\Psi_{S_xI_x}(\omega_S,\omega_I)\equiv\Psi_{SI}(\omega_S,\omega_I)H_{S_x}(\omega_S)H_{I_x}(\omega_I)$.
The integral inside the square of \eqref{eq:probability} is the heralding efficiency, or the probability that both the idler and signal photons pass through the $x^{\text{th}}$ channel of their respective WDM filters; the square is due to the two SPDC processes required.
The factor $1/4$ is the probability that the output EPR-pair is in the desired polarization Bell state.

Finally, the generated EPR-pair rate in channel $x$, reported in Fig.~\ref{fig:dist_185}, is $P_xr_{\text{P}}$, where $r_{\text{P}}=\frac{1}{10\sigma_{\text{P}}}$ is the pulse repetition rate of the pump laser.
These rates are highest in the center of the Gaussian wave-function in \eqref{eq:wavefunction} and, like the authors of \cite{shapiro2024entanglementsourcequantummemory}, we use the $B_{\text{s}}=2.430$ THz spectrum segment centered at its peak.
Employing the method from \cite[Sec.~IV-C]{shapiro2024entanglementsourcequantummemory}, we obtain 98.5\% purity of our output state, same as in Case 2 of \cite{shapiro2024entanglementsourcequantummemory}.

\end{document}